\documentclass[twocolumn,letterpaper,aps,prc,longbibliography,superscriptaddress,nofootinbib,floatfix]{revtex4-1}

\usepackage{graphicx}	

\usepackage{xspace}	

\newcommand{\pt}{\mbox{$p_T$}\xspace}

\newcommand{\Npart}{\mbox{$N_{\rm part}$}\xspace}
\newcommand{\Ncoll}{\mbox{$N_{\rm coll}$}\xspace}

\newcommand{\sqsn}{\mbox{$\sqrt{s_{_{NN}}}$}\xspace}

\newcommand{\ee}{\mbox{$e^{+}e^{-}$}\xspace}
\newcommand{\mt}{\mbox{$m_T$}\xspace}

\newcommand{\taa}{\mbox{$T_{\rm AA}$}\xspace}

\begin{document}


\title{Low-momentum direct photon measurement in Cu$+$Cu collisions at 
$\sqrt{s_{_{NN}}}=200$~GeV}

\newcommand{\abilene}{Abilene Christian University, Abilene, Texas 79699, USA}
\newcommand{\augie}{Department of Physics, Augustana University, Sioux Falls, South Dakota 57197, USA}
\newcommand{\banaras}{Department of Physics, Banaras Hindu University, Varanasi 221005, India}
\newcommand{\baruch}{Baruch College, City University of New York, New York, New York, 10010 USA}
\newcommand{\bnlcoll}{Collider-Accelerator Department, Brookhaven National Laboratory, Upton, New York 11973-5000, USA}
\newcommand{\bnlphys}{Physics Department, Brookhaven National Laboratory, Upton, New York 11973-5000, USA}
\newcommand{\caucr}{University of California-Riverside, Riverside, California 92521, USA}
\newcommand{\charlesczech}{Charles University, Ovocn\'{y} trh 5, Praha 1, 116 36, Prague, Czech Republic}
\newcommand{\chonbuk}{Chonbuk National University, Jeonju, 561-756, Korea}
\newcommand{\ciae}{Science and Technology on Nuclear Data Laboratory, China Institute of Atomic Energy, Beijing 102413, People's Republic of China}
\newcommand{\cns}{Center for Nuclear Study, Graduate School of Science, University of Tokyo, 7-3-1 Hongo, Bunkyo, Tokyo 113-0033, Japan}
\newcommand{\colorado}{University of Colorado, Boulder, Colorado 80309, USA}
\newcommand{\columbia}{Columbia University, New York, New York 10027 and Nevis Laboratories, Irvington, New York 10533, USA}
\newcommand{\czechtech}{Czech Technical University, Zikova 4, 166 36 Prague 6, Czech Republic}
\newcommand{\dapnia}{Dapnia, CEA Saclay, F-91191, Gif-sur-Yvette, France}
\newcommand{\debrecen}{Debrecen University, H-4010 Debrecen, Egyetem t{\'e}r 1, Hungary}
\newcommand{\elte}{ELTE, E{\"o}tv{\"o}s Lor{\'a}nd University, H-1117 Budapest, P{\'a}zm{\'a}ny P.~s.~1/A, Hungary}
\newcommand{\eszterhazy}{Eszterh\'azy K\'aroly University, K\'aroly R\'obert Campus, H-3200 Gy\"ongy\"os, M\'atrai \'ut 36, Hungary}
\newcommand{\fit}{Florida Institute of Technology, Melbourne, Florida 32901, USA}
\newcommand{\fsu}{Florida State University, Tallahassee, Florida 32306, USA}
\newcommand{\gsu}{Georgia State University, Atlanta, Georgia 30303, USA}
\newcommand{\hiroshima}{Hiroshima University, Kagamiyama, Higashi-Hiroshima 739-8526, Japan}
\newcommand{\howard}{Department of Physics and Astronomy, Howard University, Washington, DC 20059, USA}
\newcommand{\ihepprot}{IHEP Protvino, State Research Center of Russian Federation, Institute for High Energy Physics, Protvino, 142281, Russia}
\newcommand{\illuiuc}{University of Illinois at Urbana-Champaign, Urbana, Illinois 61801, USA}
\newcommand{\inrras}{Institute for Nuclear Research of the Russian Academy of Sciences, prospekt 60-letiya Oktyabrya 7a, Moscow 117312, Russia}
\newcommand{\instpasczech}{Institute of Physics, Academy of Sciences of the Czech Republic, Na Slovance 2, 182 21 Prague 8, Czech Republic}
\newcommand{\isu}{Iowa State University, Ames, Iowa 50011, USA}
\newcommand{\jaea}{Advanced Science Research Center, Japan Atomic Energy Agency, 2-4 Shirakata Shirane, Tokai-mura, Naka-gun, Ibaraki-ken 319-1195, Japan}
\newcommand{\jinrdubna}{Joint Institute for Nuclear Research, 141980 Dubna, Moscow Region, Russia}
\newcommand{\jyvaskyla}{Helsinki Institute of Physics and University of Jyv{\"a}skyl{\"a}, P.O.Box 35, FI-40014 Jyv{\"a}skyl{\"a}, Finland}
\newcommand{\kek}{KEK, High Energy Accelerator Research Organization, Tsukuba, Ibaraki 305-0801, Japan}
\newcommand{\korea}{Korea University, Seoul, 136-701, Korea}
\newcommand{\kurchatov}{National Research Center ``Kurchatov Institute", Moscow, 123098 Russia}
\newcommand{\kyoto}{Kyoto University, Kyoto 606-8502, Japan}
\newcommand{\labllr}{Laboratoire Leprince-Ringuet, Ecole Polytechnique, CNRS-IN2P3, Route de Saclay, F-91128, Palaiseau, France}
\newcommand{\lahorelums}{Physics Department, Lahore University of Management Sciences, Lahore 54792, Pakistan}
\newcommand{\lawllnl}{Lawrence Livermore National Laboratory, Livermore, California 94550, USA}
\newcommand{\losalamos}{Los Alamos National Laboratory, Los Alamos, New Mexico 87545, USA}
\newcommand{\lpc}{LPC, Universit{\'e} Blaise Pascal, CNRS-IN2P3, Clermont-Fd, 63177 Aubiere Cedex, France}
\newcommand{\lund}{Department of Physics, Lund University, Box 118, SE-221 00 Lund, Sweden}
\newcommand{\lyon}{IPNL, CNRS/IN2P3, Univ Lyon, Université Lyon 1, F-69622, Villeurbanne, France}
\newcommand{\maryland}{University of Maryland, College Park, Maryland 20742, USA}
\newcommand{\mass}{Department of Physics, University of Massachusetts, Amherst, Massachusetts 01003-9337, USA}
\newcommand{\michigan}{Department of Physics, University of Michigan, Ann Arbor, Michigan 48109-1040, USA}
\newcommand{\muenster}{Institut f\"ur Kernphysik, University of M\"unster, D-48149 M\"unster, Germany}
\newcommand{\myongji}{Myongji University, Yongin, Kyonggido 449-728, Korea}
\newcommand{\nagasaki}{Nagasaki Institute of Applied Science, Nagasaki-shi, Nagasaki 851-0193, Japan}
\newcommand{\nara}{Nara Women's University, Kita-uoya Nishi-machi Nara 630-8506, Japan}
\newcommand{\natmephi}{National Research Nuclear University, MEPhI, Moscow Engineering Physics Institute, Moscow, 115409, Russia}
\newcommand{\newmex}{University of New Mexico, Albuquerque, New Mexico 87131, USA}
\newcommand{\nmsu}{New Mexico State University, Las Cruces, New Mexico 88003, USA}
\newcommand{\ohio}{Department of Physics and Astronomy, Ohio University, Athens, Ohio 45701, USA}
\newcommand{\ornl}{Oak Ridge National Laboratory, Oak Ridge, Tennessee 37831, USA}
\newcommand{\orsay}{IPN-Orsay, Univ.~Paris-Sud, CNRS/IN2P3, Universit\'e Paris-Saclay, BP1, F-91406, Orsay, France}
\newcommand{\peking}{Peking University, Beijing 100871, People's Republic of China}
\newcommand{\pnpi}{PNPI, Petersburg Nuclear Physics Institute, Gatchina, Leningrad region, 188300, Russia}
\newcommand{\riken}{RIKEN Nishina Center for Accelerator-Based Science, Wako, Saitama 351-0198, Japan}
\newcommand{\rikjrbrc}{RIKEN BNL Research Center, Brookhaven National Laboratory, Upton, New York 11973-5000, USA}
\newcommand{\rikkyo}{Physics Department, Rikkyo University, 3-34-1 Nishi-Ikebukuro, Toshima, Tokyo 171-8501, Japan}
\newcommand{\saispbstu}{Saint Petersburg State Polytechnic University, St.~Petersburg, 195251 Russia}
\newcommand{\saopaulo}{Universidade de S{\~a}o Paulo, Instituto de F\'{\i}sica, Caixa Postal 66318, S{\~a}o Paulo CEP05315-970, Brazil}
\newcommand{\seoulnat}{Department of Physics and Astronomy, Seoul National University, Seoul 151-742, Korea}
\newcommand{\stonybrkc}{Chemistry Department, Stony Brook University, SUNY, Stony Brook, New York 11794-3400, USA}
\newcommand{\stonycrkp}{Department of Physics and Astronomy, Stony Brook University, SUNY, Stony Brook, New York 11794-3800, USA}
\newcommand{\subatech}{SUBATECH (Ecole des Mines de Nantes, CNRS-IN2P3, Universit{\'e} de Nantes) BP 20722-44307, Nantes, France}
\newcommand{\tenn}{University of Tennessee, Knoxville, Tennessee 37996, USA}
\newcommand{\titech}{Department of Physics, Tokyo Institute of Technology, Oh-okayama, Meguro, Tokyo 152-8551, Japan}
\newcommand{\tsukuba}{Tomonaga Center for the History of the Universe, University of Tsukuba, Tsukuba, Ibaraki 305, Japan}
\newcommand{\vandy}{Vanderbilt University, Nashville, Tennessee 37235, USA}
\newcommand{\waseda}{Waseda University, Advanced Research Institute for Science and Engineering, 17  Kikui-cho, Shinjuku-ku, Tokyo 162-0044, Japan}
\newcommand{\weizmann}{Weizmann Institute, Rehovot 76100, Israel}
\newcommand{\wigner}{Institute for Particle and Nuclear Physics, Wigner Research Centre for Physics, Hungarian Academy of Sciences (Wigner RCP, RMKI) H-1525 Budapest 114, POBox 49, Budapest, Hungary}
\newcommand{\yonsei}{Yonsei University, IPAP, Seoul 120-749, Korea}
\affiliation{\abilene}
\affiliation{\augie}
\affiliation{\banaras}
\affiliation{\baruch}
\affiliation{\bnlcoll}
\affiliation{\bnlphys}
\affiliation{\caucr}
\affiliation{\charlesczech}
\affiliation{\chonbuk}
\affiliation{\ciae}
\affiliation{\cns}
\affiliation{\colorado}
\affiliation{\columbia}
\affiliation{\czechtech}
\affiliation{\dapnia}
\affiliation{\debrecen}
\affiliation{\elte}
\affiliation{\eszterhazy}
\affiliation{\fit}
\affiliation{\fsu}
\affiliation{\gsu}
\affiliation{\hiroshima}
\affiliation{\howard}
\affiliation{\ihepprot}
\affiliation{\illuiuc}
\affiliation{\inrras}
\affiliation{\instpasczech}
\affiliation{\isu}
\affiliation{\jaea}
\affiliation{\jinrdubna}
\affiliation{\jyvaskyla}
\affiliation{\kek}
\affiliation{\korea}
\affiliation{\kurchatov}
\affiliation{\kyoto}
\affiliation{\labllr}
\affiliation{\lahorelums}
\affiliation{\lawllnl}
\affiliation{\losalamos}
\affiliation{\lpc}
\affiliation{\lund}
\affiliation{\lyon}
\affiliation{\maryland}
\affiliation{\mass}
\affiliation{\michigan}
\affiliation{\muenster}
\affiliation{\myongji}
\affiliation{\nagasaki}
\affiliation{\nara}
\affiliation{\natmephi}
\affiliation{\newmex}
\affiliation{\nmsu}
\affiliation{\ohio}
\affiliation{\ornl}
\affiliation{\orsay}
\affiliation{\peking}
\affiliation{\pnpi}
\affiliation{\riken}
\affiliation{\rikjrbrc}
\affiliation{\rikkyo}
\affiliation{\saispbstu}
\affiliation{\saopaulo}
\affiliation{\seoulnat}
\affiliation{\stonybrkc}
\affiliation{\stonycrkp}
\affiliation{\subatech}
\affiliation{\tenn}
\affiliation{\titech}
\affiliation{\tsukuba}
\affiliation{\vandy}
\affiliation{\waseda}
\affiliation{\weizmann}
\affiliation{\wigner}
\affiliation{\yonsei}
\author{A.~Adare} \affiliation{\colorado} 
\author{S.~Afanasiev} \affiliation{\jinrdubna} 
\author{C.~Aidala} \affiliation{\columbia} \affiliation{\michigan} 
\author{N.N.~Ajitanand} \altaffiliation{Deceased} \affiliation{\stonybrkc} 
\author{Y.~Akiba} \email[PHENIX Spokesperson: ]{akiba@rcf.rhic.bnl.gov} \affiliation{\riken} \affiliation{\rikjrbrc} 
\author{H.~Al-Bataineh} \affiliation{\nmsu} 
\author{J.~Alexander} \affiliation{\stonybrkc} 
\author{M.~Alfred} \affiliation{\howard} 
\author{K.~Aoki} \affiliation{\kek} \affiliation{\kyoto} \affiliation{\riken} 
\author{L.~Aphecetche} \affiliation{\subatech} 
% ...
\collaboration{PHENIX Collaboration} \noaffiliation

\date{\today}


\begin{abstract}


We have measured direct photons for $p_T<5~$GeV/$c$ in minimum bias and 
0\%--40\% most central events at midrapidity for Cu$+$Cu collisions at 
$\sqrt{s_{_{NN}}}=200$ GeV.  The $e^{+}e^{-}$ contribution from 
quasi-real direct virtual photons has been determined as an excess over 
the known hadronic contributions in the $e^{+}e^{-}$ mass distribution. 
A clear enhancement of photons over the binary scaled $p$$+$$p$ fit is 
observed for $p_T<4$ GeV/$c$ in Cu$+$Cu data. The $p_T$ spectra are 
consistent with the Au$+$Au data covering a similar number of 
participants. The inverse slopes of the exponential fits 
to the excess after subtraction of the $p$$+$$p$ baseline are
285$\pm$53(stat)$\pm$57(syst)~MeV/$c$ and 
333$\pm$72(stat)$\pm$45(syst)~MeV/$c$ for minimum bias and 0\%--40\% 
most central events, respectively. The rapidity density, $dN/dy$, of 
photons demonstrates the same power law as a function of $dN_{\rm ch}/d\eta$ 
observed in Au$+$Au at the same collision energy.

\end{abstract}

\maketitle

\section{Introduction}

Direct photons are excellent probes for understanding the time evolution 
of the hot and dense matter created in ultrarelativistic heavy ion 
collisions~\cite{Adcox:2004mh,Adams:2005dq}.  Direct photons are 
produced throughout the collision and carry information about the 
medium at the time when the photons were emitted, because the only 
interaction is electromagnetic~\cite{David:2006sr}. Direct photons are 
produced via interactions at partonic and hadronic levels in either 
initial hard scatterings of the collision or thermal radiation from the 
medium and, by definition, do not originate from hadron 
decays~\cite{Stankus:2005eq}.  In particular, thermal photons, which 
contribute dominantly at low momentum~\cite{Turbide:2003si}, are one of 
the most important probes because they allow us direct access to the 
thermodynamic properties of the created medium. However, photons from 
hadron decays account for a large fraction in the inclusive photon 
yield, typically more than 80\% for heavy ion collisions. The large 
number of decay photons makes the measurement challenging.

Two analysis methods, the virtual photon method~\cite{Adare:2008ab} and 
the external conversion method~\cite{Adare:2014fwh}, have been 
established to measure direct photons at low \pt ($\pt<5~$GeV/$c$). 
Low-\pt direct photon measurements have been made in PHENIX and STAR 
experiments at the Relativistic Heavy Ion Collider (RHIC) for not only 
in Au$+$Au collisions~\cite{Adare:2008ab,Adare:2014fwh,STAR:2016use} but 
also in $p$$+$$p$ and $d$$+$Au~\cite{Adare:2012vn} collisions. The virtual 
photon method makes it possible to measure direct photons even if the 
signal to background is only a few percent as in $p$$+$$p$ and $d$$+$Au 
collisions, while in Au$+$Au collisions S/B reaches 15\%. The $p$$+$$p$ 
measurement allows to determine the hard photon yield from initial hard 
scatterings. No significant modification of the \pt distribution of 
direct photons due to cold nuclear effects is seen in the $d$$+$Au data. 
Finally, an enhanced yield of low-\pt direct photons, which is 
unexplainable by hard photon production and cold nuclear matter effects, 
has been discovered in Au$+$Au collisions in central and semi-central 
events at $\sqsn=200$~GeV~\cite{Adare:2008ab,Adare:2014fwh}.

The ALICE experiment at the Large Hadron Collider (LHC) has also 
succeeded in measuring the low-\pt direct photons with the external 
conversion method in Pb$+$Pb collisions at 
$\sqsn=2.76$~TeV~\cite{Adam:2015lda} and observed a larger yield and a 
higher inverse slope of the spectrum than at RHIC, implying that a 
larger and hotter thermalized medium is produced at the LHC energy. 
Further understanding of the thermal properties of the created hot 
medium can be realized through the systematic study of low-\pt direct 
photon production within a wide range of system size and collision 
energy.

In this paper, we present the measurement of low-\pt direct photons in 
Cu$+$Cu collisions at $\sqsn=200$~GeV with the virtual photon method. This 
measurement may provide additional information on the system size 
dependence of low-\pt direct photon production. This paper focuses on 
two centrality classes, minimum bias (MB) and 0\%--40\% most central 
collisions, for which the number of participants, $N_{\rm part}$, is similar 
to peripheral Au$+$Au ($N_{\rm part}=34.6\pm1.2$~\cite{Adare:2013yxp} and 
66.4$\pm$2.5~\cite{Adare:2015bua}).

\section{The PHENIX detector}

\begin{figure}[thb]
\includegraphics[width=1.0\linewidth]{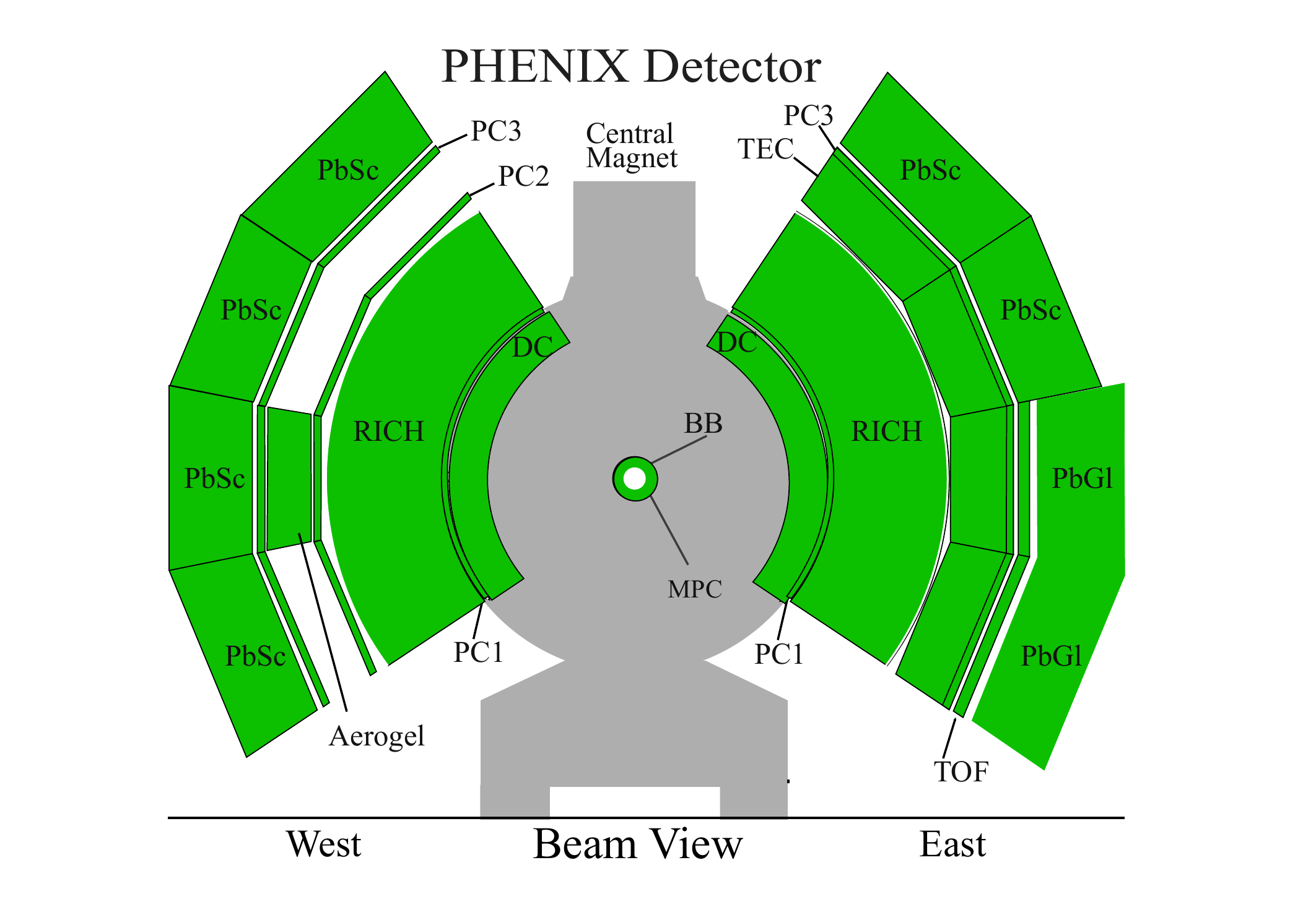}
\caption{\label{fig:phenix_detctor}
The beam view of the PHENIX detector configuration in 2005.}
\end{figure}

The two PHENIX central arm spectrometers in conjunction with the 
beam-beam counters (BBC) are used for this measurement.
Figure~\ref{fig:phenix_detctor} shows the beam view of the PHENIX 
detector configuration for the 2005 run. The BBCs, with rapidity 
coverage $3.1<|\eta|<3.9$, are located at $\pm144~$cm away from the 
nominal interaction point. They measure charged particles that are used 
to determine the $z$-vertex position, centrality, and the event plane. 
They provide the MB event trigger with a trigger 
efficiency of 94\%. The two central arms cover $|\eta|<0.35$ and an 
azimuthal angle range of $\pi/2$ per arm. Each arm is instrumented with 
a drift chamber (DC) and pad chambers (PCs) that determine the 
trajectories and, together with a magnetic field, measure the momenta of 
charged particles. The material in front of the DC is minimal, 0.39\% of 
a radiation length, to allow for a good momentum resolution of $\delta 
p/p=1\% \oplus 1.1\% \times p$~[GeV/$c$] above 
0.2~GeV/$c$~\cite{PHENIX-NIM}, and to minimize the amount of photon 
conversions. Eight separate sectors of electromagnetic calorimeters 
(EMCals) composed of two lead-glass (PbGl) calorimeters in the bottom 
sectors of the east arm and six lead-scintillator (PbSc) calorimeters 
for the remainder, provide an electromagnetic shower energy measurement 
with resolution $\Delta E/E$ of $2.1\%\oplus8.1\%/\sqrt{E}$ for PbSc and 
of $0.8\%\oplus5.9\%/\sqrt{E}$ for PbGl ($E$ in GeV)~\cite{PHENIX-NIM}. 
Requiring energy-momentum matching with an associated hit in the Ring 
Imaging \v{C}erenkov counter (RICH) provides a hadron rejection factor 
of better than 10$^{4}$, thus providing good electron identification. 
The mass resolution for $\ee$ pairs is determined with a Monte Carlo 
simulation which is tuned to match the shape of the reconstructed $\ee$ 
mass distribution in the data below 90~MeV/$c^2$~\cite{Adare:2014mgk}, 
where $\ee$ pairs from $\pi^0$ Dalitz decays are dominant. The 
calculated $\ee$ mass resolution is $\sigma_{ee} = 3.1~$MeV/$c^2$ for 
$1<\pt<2~$GeV/$c$, and it increases by about 1~MeV/$c^2$ per GeV/$c$ as 
\pt increases.

\section{Analysis}

Low-\pt direct photons, measured using the virtual photon method, are 
the subject of this analysis. Any production process of direct photons 
has a higher order process producing a quasi-real virtual photon, which 
then produces a low mass, high \pt $\ee$ pair. The relation between 
the photon emission ($dN_{\gamma}$) and associate electron pair rates 
($dN_{ee}$) is expressed as:

\begin{equation}
\frac{d^{2}N_{ee}}{dm_{ee}}=\frac{2\alpha}{3\pi}\frac{1}{m_{ee}}\sqrt{1-\frac{4m_{e}^{2}}{m_{ee}^{2}}}
  (1+\frac{2m_{e}^{2}}{m_{ee}^{2}})SdN_{\gamma},
\label{eq:sfactor}
\end{equation}

\noindent where $\alpha, m_{e}, m_{ee}$ are the fine structure constant 
and masses for the electron and the electron pair. $S$ is introduced to 
factor out the difference between real and virtual photon emission. It 
is a process dependent factor because it accounts for the effects of 
form factors, phase space and spectral functions~\cite{Adare:2009qk}. 
For direct virtual photons satisfying $\pt\gg m_{ee}$, $S$ is almost 
unity, while it drops to 0 as $m_{ee}$ approaches the parent hadron mass 
in case of hadron decays. As a result, $S$ introduces a shape difference 
of the $\ee$ mass distributions for virtual photons and hadron decays. 
The key idea of this measurement is to utilize this shape difference. 
Therefore, the contribution of the $\ee$ pairs internally converted via 
virtual photons is determined as an excess yield over the known hadronic 
contributions in the mass region above the $\pi^{0}$ mass, typically 
$0.1<m_{ee}<0.3~$GeV/$c^{2}$, by a template fit. The direct photon 
fraction at $m_{ee}=0$ is then obtained by extrapolation of the template 
fit result. Finally, the obtained direct photon fraction can be 
converted to the real direct photon yield using the measured inclusive 
photon yield. A detailed description of the virtual photon method can be 
found in Ref.~\cite{Adare:2009qk}.

This measurement is based on a MB sample of $4.95\times10^{8}$ 200~GeV 
Cu$+$Cu collisions with $z$-vertex within $25~$cm of the nominal 
interaction point collected in 2005, equivalent to 0.44~nb$^{-1}$. All 
electrons with $p^{e}_{T}>0.3~$GeV/$c$ are paired in each event. These 
$\ee$ pairs are required to have $\pt>1~$GeV/$c$.

\begin{figure}[thb]
\includegraphics[width=1.0\linewidth]{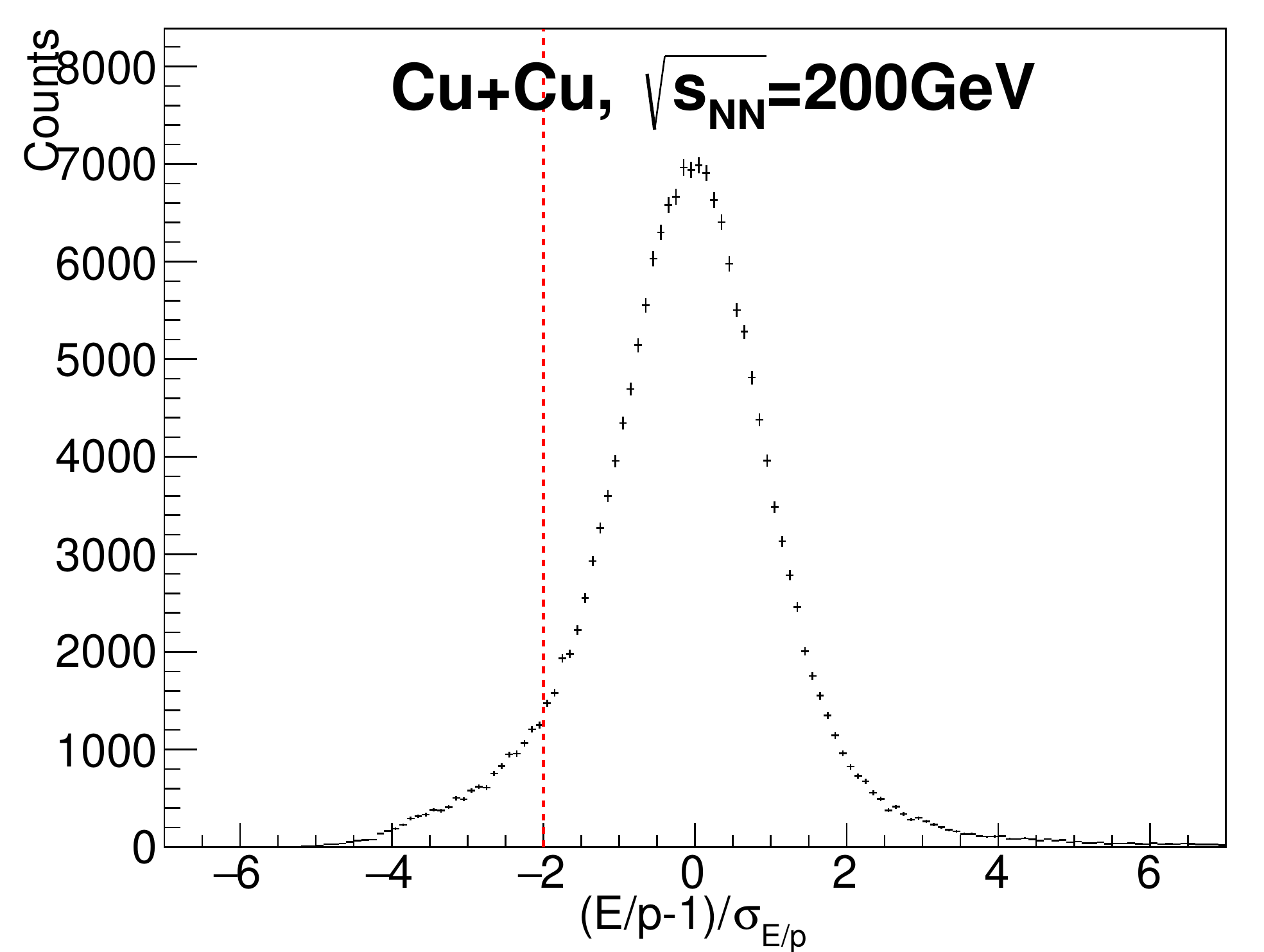}
\caption{\label{fig:dep_dist} 
The $E/p$ distribution for electrons from pairs with \pt of 1--2~GeV/$c$ 
after applying all cuts for electron identification except for $E/p$.}
\end{figure}

Figure~\ref{fig:dep_dist} shows the $E/p$ distribution for electrons 
from pairs with \pt of 1--2~GeV/$c$, where $E$ is measured with the 
EMCal, and $p$ from the track radius in the magnetic field. All electron 
identification cuts except for $E/p$ are applied for this figure. Because 
hadrons do not deposit their full energy in the EMCal, hadron 
contamination produces a tail in the negative region.  This plot 
indicates the excellent purity of the electron sample. All electron 
candidates are required to have $(E/p-1)/\sigma_{E/p}>-2$, resulting in 
negligible hadron contamination.

Undesired pairs from several background sources contaminate the 
foreground pair distribution. The first source is fake pairs due to 
accidentally overlapping hits in various detectors. RICH ring-sharing 
and cluster overlaps in the PCs are the main sources for these fake 
pairs. They can be removed by geometric analysis 
cuts~\cite{Adare:2009qk,Adare:2015ila}. The RICH ring-sharing cut 
requires separation of ring centers for the two electrons of a pair to 
be greater than 25~cm, which is larger than the expected maximum 
diameter of a RICH ring, $\sim16.8~$cm. Tracks are also required to be 
separated by $\Delta z>0.5~$cm and $\Delta\phi>0.02~$rad to remove 
overlap in the PCs.

The second background source is photon conversions in the detector 
material. These can be eliminated because the PHENIX tracking algorithm, 
which assumes all tracks come from the collision vertex, introduces an 
artificial opening angle of the conversion pairs with the decay plane 
perpendicular to the magnetic field.

\subsection{Background evaluation}\label{sec:bg_eval}

After removing the detector-oriented fake pairs and conversions, the 
foreground distributions for unlike-sign (FG$_{+-}$) and like-sign pairs 
(FG$_{--}$, FG$_{++}$) can be expressed as:

\begin{eqnarray}
{\rm FG}_{--} & = &     {\rm BG}^{\rm CM}_{--} 
+ {\rm BG}^{\rm JT}_{--} + {\rm BG}^{\rm XC}_{--} = 
{\rm BG}^{\rm SUM}_{--}, \\ 
{\rm FG}_{++} & = &     {\rm BG}^{\rm CM}_{++} 
+ {\rm BG}^{\rm JT}_{++} + {\rm BG}^{\rm XC}_{++} = 
{\rm BG}^{\rm SUM}_{++}, \\ 
{\rm FG}_{+-}     & = & S + {\rm BG}^{\rm SUM}_{+-} + {\rm HD}_{+-}. 
\label{eq:unlike_fg}
\end{eqnarray}

\noindent Here FG refers to the data and BG refers to backgrounds whose 
shapes are calculated as described below, but whose normalization comes 
from a fit to the data (FG). $S$ refers to the direct virtual photon 
signal and HD refers to correlated pairs from known hadron decays. It is 
notable that the like-sign pair distributions are composed of only 
random combinations (BG$^{\rm CM}$), jet-induced correlations (BG$^{\rm 
JT}$) and correlated fake pairs from double Dalitz decays of the 
$\pi^{0}, \eta$ (BG$^{\rm XC}$).  The sum of these backgrounds is 
referred to as BG$^{\rm SUM}$ in this paper. Once compositions of these 
background contributions are known in the like-sign combination sample, 
the unlike-sign combination background, BG$^{\rm SUM}_{+/-}$, can be 
determined within the same analysis framework.

\subsubsection{Combinatorial background}

The combinatorial background can be reproduced by the event mixing 
technique with event classification with respect to $z$-vertex position, 
event plane, and centrality. However the modulation of the mass 
distribution by the elliptic flow, which is apparent in the real events, 
is not fully introduced in event mixing because of the limited reaction 
plane resolution. Thus, pairs in the mixed events are weighted by a 
factor based on the measured azimuthal anisotropy of single 
electrons~\cite{Adare:2015ila} for given reaction plane classes. The 
weighting factor, $w$, depending on the opening angle of a pair is 
calculated as:

\begin{equation}
w(\Delta\phi) = 1 + 2v_{2}^{a}v_{2}^{b}\cos2(\Delta\phi),
\label{eq:flow_mod}
\end{equation}

\noindent where $\Delta\phi, v_{2}^{a,b}$ are the pair opening angle and 
azimuthal anisotropy of each electron in a pair, respectively. The flow 
modulation makes at most a few percent difference in the mass shape.

\subsubsection{Jet-induced correlation}

Jet-induced correlations are pairs in which each electron is from a 
different parent, but both parents are from the same jet or back-to-back 
jets. Such events are simulated by {\sc pythia8}~\cite{Sjostrand:2006za, 
Sjostrand:2007gs} with {\sc cteq5l}~\cite{Lai:1999wy} parton 
distribution functions. The {\sc pythia8}-generated events are passed 
through a {\sc geant3}~\cite{Brun:1987ma} based simulation of the PHENIX 
detector in which all detector effects such as the acceptance and 
efficiencies are taken into account. Uncorrelated combinations are 
evaluated by the event mixing technique within the simulated events. It 
is found that the shape of the like-sign mass distribution for the 
uncorrelated combinations is consistent with that for the foreground 
combinations in $0.6<m_{ee}<1.1~$GeV/$c^2$. Here, the true and other 
correlated pairs are removed from the foreground distribution before the 
comparison. Normalization of the uncorrelated combinations in a specific 
region of a pair opening angle, where opening angle distributions for 
correlated and uncorrelated pairs are consistent, gives a consistent 
result.  A detailed description can be found in Ref.~\cite{Adare:2015ila}. 
Finally the jet-induced correlations are obtained by removing 
uncorrelated combinations from the simulated mass distribution.

\subsubsection{Correlated Dalitz and Double Dalitz Cross Pairs}

The other nonnegligible source of correlated background is cross 
combinations from decays having two electron pairs in the final state, 
i.e. $\pi^{0}$ and $\eta$ double Dalitz decays and Dalitz decays with a 
subsequent photon conversion. These cross combinations are localized at 
the very low mass region below the $\pi^{0}$ and $\eta$ masses. The mass 
distributions of these cross combinations from $\pi^{0}$ and $\eta$ are 
calculated using the aforementioned {\sc geant3} simulation with the $\pi^{0}$ 
and $\eta$ distributions measured by PHENIX.

\subsubsection{Background Normalization by BG$^{\rm SUM}$ fit}

\begin{figure}[thb]
\includegraphics[width=1.0\linewidth]{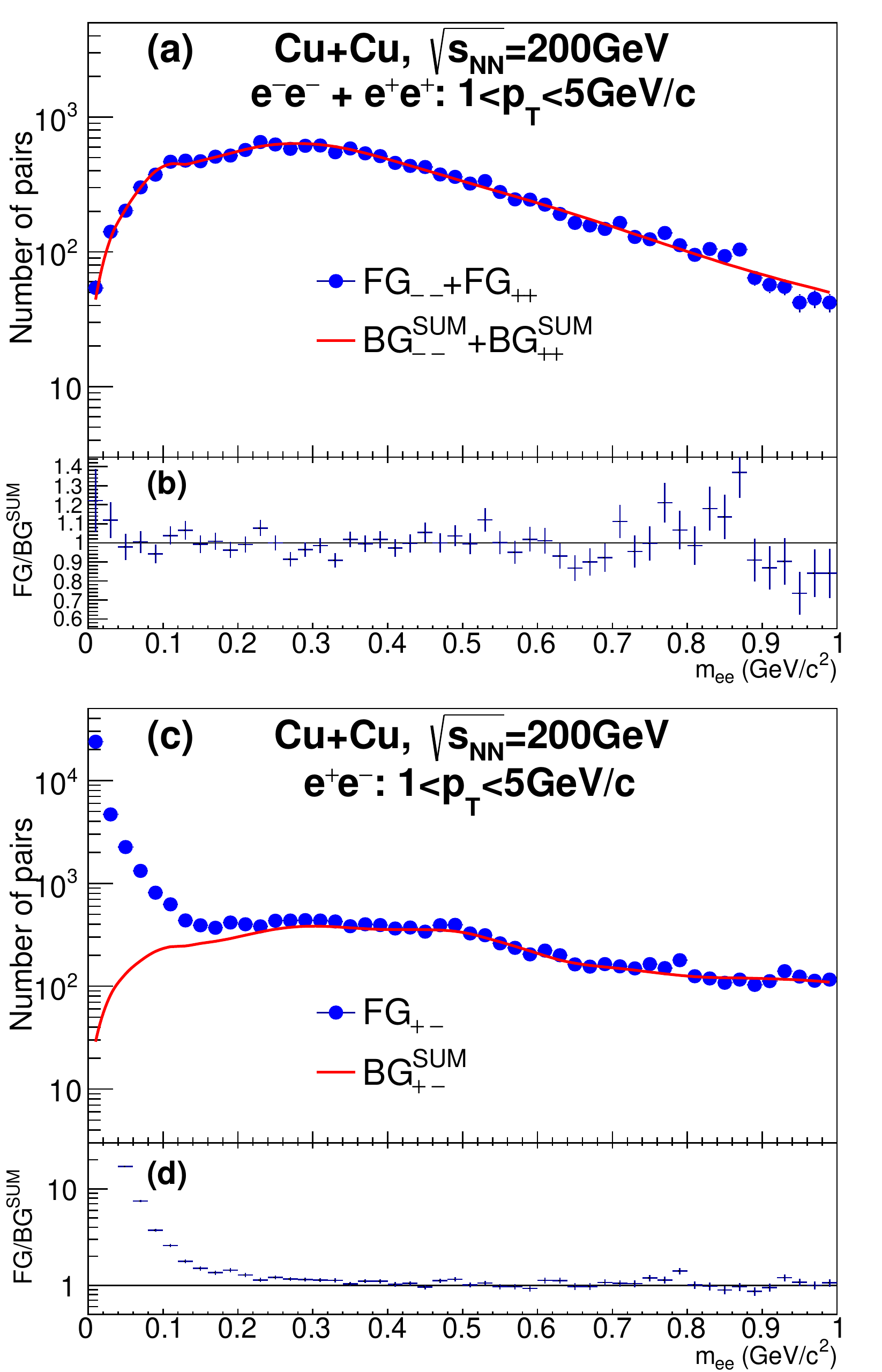}
\caption{\label{fig:comp_bg}
The (a) like-sign and (c) unlike-sign mass distributions of the data together 
with BG$^{\rm SUM}$ normalized by the BG$^{\rm SUM}$ fit for $1<\pt<5~$GeV/$c$. 
(b) and (d) the ratios of data over BG$^{\rm SUM}$.}
\end{figure}

The calculated BG$_{--,++}$ distributions are the ingredients for a fit 
to FG$_{--,++}$, which then yields the contribution of each component 
to the background, BG$^{\rm SUM}$. Pairs from the same jet and back-to-back 
jets are separately included in the fit because they are influenced 
differently by jet quenching. The BG$^{\rm SUM}$ fit to FG$_{--,++}$ works 
very well. Figure~\ref{fig:comp_bg} shows the like-sign and unlike-sign 
mass distributions of the data together with BG$^{\rm SUM}$ normalized by 
the BG$^{\rm SUM}$ fit for $1<\pt<5~$GeV/$c$ where the virtual photon 
analysis is performed. The normalized BG$^{\rm SUM}$ is in good agreement 
with the data for like-sign pairs. The contribution of the physically 
correlated pairs ($S+{\rm HD}_{+-}$ in Eq.~\ref{eq:unlike_fg}) is significant 
in the foreground unlike-sign pair mass distribution below 
0.3~GeV/$c^2$.

A cross check with the like-sign subtraction method~\cite{Adare:2017caq} 
is done to demonstrate that the BG$^{\rm SUM}_{+-}$ properly accounts for all 
backgrounds. To infer the background in unlike-sign distributions, a 
correction must be made to account for the relative acceptance 
difference between like- and unlike-sign pairs. Thus, the 
acceptance-corrected like-sign pairs should be expressed as:

\begin{equation}
{\rm BG}^{\rm SUM}_{+-} = \alpha_{\rm acc} \times ({\rm FG}_{--}+{\rm FG}_{++}).
\label{eq:flow_mod2}
\end{equation} 
The acceptance correction factor, $\alpha_{\rm acc}$, is calculated as the 
ratio of like- and unlike-sign pairs from mixed events.

\begin{figure*}[thb]
\includegraphics[width=0.99\linewidth]{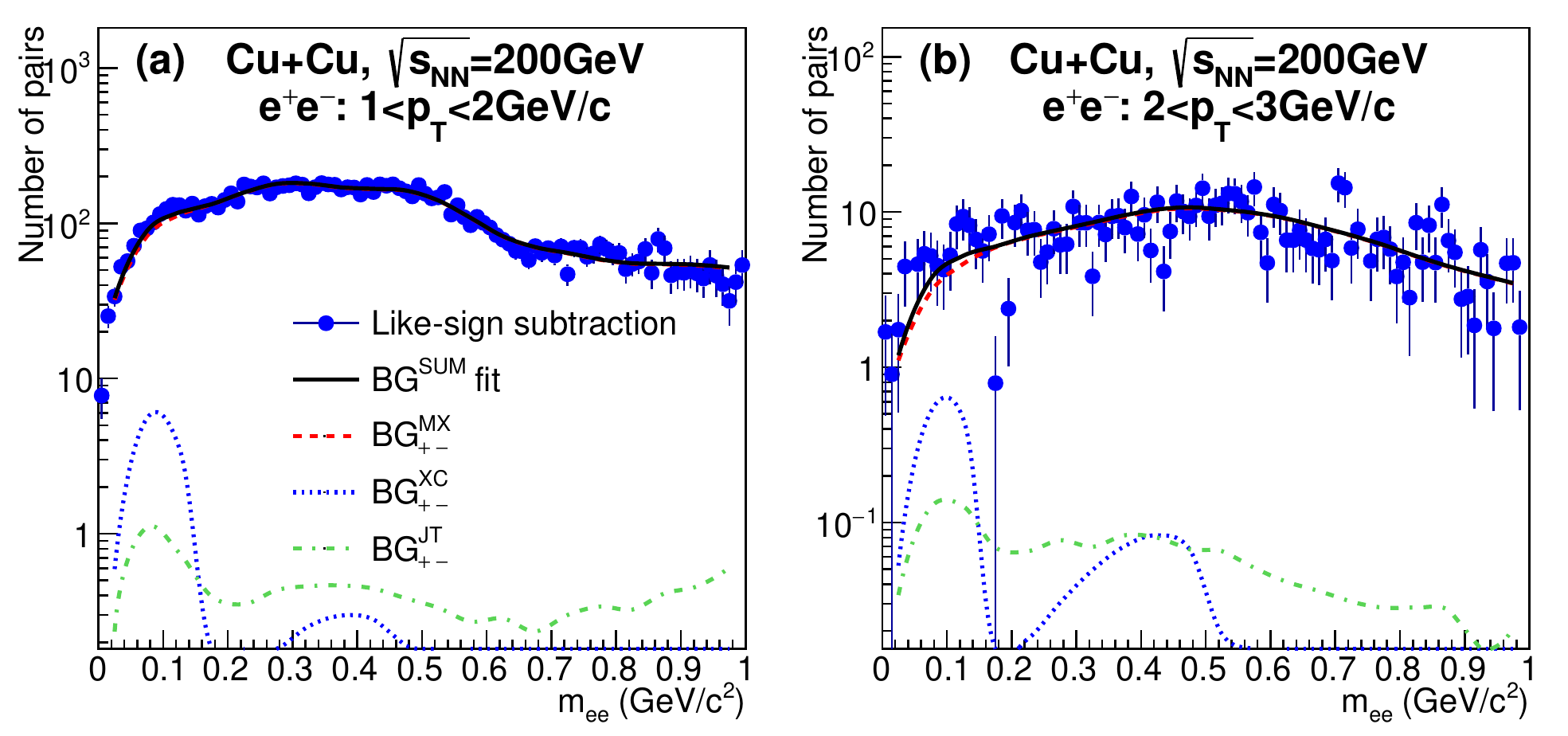}
\caption{\label{fig:bg_xcheck}
The background pair distributions of $\ee$ determined by the like-sign 
subtraction method, $\alpha_{\rm acc}\times($FG$_{--}+$FG$_{++})$, 
(circle symbols) and BG$^{\rm SUM}$ fit method (solid curves) for (a) 
$\pt=$1--2~GeV/$c$ and (b) 2--3~GeV/$c$. The resulting contributions to 
BG$^{\rm SUM}_{+-}$ are also shown by dashed, dotted, and dashed-dotted 
curves [see text and legend].}
\end{figure*}

Figure~\ref{fig:bg_xcheck} shows the background pair distributions of 
$\ee$ determined by the like-sign subtraction technique and the method 
used here for $p_{T}$ of 1--2 and 2--3~GeV/$c$, respectively. The two 
distributions are consistent within the statistical errors. The present 
method yields a smaller uncertainty, particularly at high $p_{T}$. The 
combinatorial background (dashed [red] curves) has a much more 
significant contribution in BG$^{\rm SUM}$ compared to those of the 
cross pairs (dotted [blue] curves) and jet-induced correlations 
(dashed-dotted [green] curves).

\subsubsection{Correlated pairs from hadron decays}

The last $\ee$ background source (indicated as $HD_{+-}$) for the direct 
virtual photon signal is the known hadron decays. The invariant yields 
of $\pi^{0}$ in the 200~GeV Cu$+$Cu as measured by 
PHENIX~\cite{Adare:2008ad} have been successfully parameterized by a 
modified Hagedorn fit:

\begin{equation}
E\frac{d^{3}\sigma}{dp^{3}} = A (e^{-(a p_{T}+b p_{T}^{2})}+p_{T}/p_{0})^{-n}.
\label{eq:mod_Hagedorn}
\end{equation}
The resulting Hagedorn fit parameters for 0\%--40\% and MB samples are 
listed in Table~\ref{tb:fit_par_Hagedorn}.

\begin{table}[tbh]
\caption{Hagedorn fit parameters for the $\pi^{0}$ distribution in 
0\%--40\% centrality and MB in Cu$+$Cu collisions.}
\label{tb:fit_par_Hagedorn}
\begin{ruledtabular} \begin{tabular}{ccc}
Fit parameter               & 0\%--40\%                        & MB \\ 
\hline
$A$ [mb GeV$^{-2}c^{3}$]    & (3.5$\pm$2.8)$\times10^{2}$   & (1.8$\pm$0.6)$\times10^{2}$ \\
$a$ [(GeV$/c$)$^{-1}$]      & 0.41$\pm$0.22                 & 0.42$\pm$0.09 \\ 
$b$ [(GeV$/c$)$^{-2}$]      & 0.22$\pm$0.16                 & 0.20$\pm$0.07 \\
$p_{0}$ [GeV$/c$]           & 0.70$\pm$0.09                 & 0.69$\pm$0.04 \\
$n$                         & 8.02$\pm$0.15                 & 8.01$\pm$0.07 \\
\end{tabular} \end{ruledtabular}
\end{table}

Note that the large uncertainty of the absolute scale parameter, $A$, 
does not affect the direct photon result because only the shape enters 
in determining the direct photon fraction. A detailed description of 
this analysis appears in the next section, Sec.~\ref{sec:det_r_gamma}. 
$\mt$-scaling of the parameterized $\pi^{0}$ yield has been shown to 
accurately reproduce the invariant yields of other known 
hadrons~\cite{Adare:2013yxp}. All known hadron decays producing $\ee$ 
are simulated with this parameterization by a Monte Carlo event 
generator within the PHENIX framework~\cite{Adare:2009qk} and passed 
through the PHENIX {\sc geant3} simulation. The simulated $\ee$ pair mass 
distributions for known hadrons are merged as a ``cocktail'' of the 
hadron decay contributions. The particle compositions in the hadronic 
cocktail are based on the measured yields. The particle ratios to the 
$\pi^0$ yield are identical to the $p$$+$$p$ data~\cite{Adler:2006hu}.

An additional source of decay background is $\ee$ pairs from open heavy 
flavor decays. They hide behind the cocktail of photonic decays 
discussed previously in the mass region of interest below 
$m_{ee}$=0.3~GeV/$c^{2}$.  Their contribution becomes significant only 
around 0.6~GeV/$c^{2}$, and then dominant in the high mass region above 
1~GeV/$c^{2}$ because of their large opening angle. Their low mass 
contribution can be extrapolated using a model fit to the data in the 
high mass region~\cite{Adare:2017caq}. PHENIX has reported that the low 
mass distribution has a model dependence~\cite{Adare:2015ila}. This 
model dependence results in a 100\% uncertainty particularly on the 
$c\bar{c}$ contribution. The open heavy flavor contribution is evaluated 
by binary-scaling of the $d$$+$Au result~\cite{Adare:2017caq}. However, 
the $c\bar{c}$ contribution is less than 0.1\% at most in the mass 
region of interest, $0.3~$GeV/$c^2$, even if 100\% uncertainty from the 
model dependence is taken into account.

\subsection{Determination of direct photon fraction}\label{sec:det_r_gamma}

The direct virtual photon signal is now extracted as the remainder of 
the signal above the backgrounds described in the previous section, 
Sec.~\ref{sec:bg_eval}. A similar fitting procedure to the one described 
in Ref.~\cite{Adare:2008ab} is employed, in which Eq.~\ref{eq:fit_gamma} 
is fit to the mass distribution, with the following difference. In the 
previous analysis only the hadronic cocktail was included in the fit. In 
the present measurement, the open heavy flavor and BG$^{\rm SUM}$ 
contributions, which were subtracted before the fit in the previous 
measurements, are now included together with the hadronic cocktail as 
fixed contributions in the fit as Eq.~\ref{eq:fit_gamma}.  This is done 
in order for a log-likelihood fit to work properly even with limited 
statistics in the data, especially at higher \pt.

\begin{equation}
f(m_{ee}) = (1-r_{\gamma})f_{c}(m_{ee}) + r_{\gamma}f_{\rm dir}(m_{ee}) + f_{\rm BG}(m_{ee}),
\label{eq:fit_gamma}
\end{equation}

\noindent where $r_{\gamma}$ is the only fit parameter and $f_c, f_{\rm BG}$ 
are the hadronic cocktail and the fixed contribution of a sum of the open 
heavy flavor and BG$^{\rm SUM}$ pairs, respectively. The expected mass shape 
of the direct virtual photons, $f_{\rm dir}$, is calculated by a Monte Carlo 
simulation based on Eq.~\ref{eq:sfactor}.  It does not show the drop 
that appears in the mass shapes of $\ee$ pairs from $\pi^0, \eta$ Dalitz 
decays because of $S\sim1$ in Eq.~\ref{eq:sfactor}. $f_{\rm dir}, f_{c}$ are 
normalized for $m_{ee}<0.03~$GeV/$c^{2}$ before the fit to ensure the 
fit result matches the data at $m_{ee}=0$, where $f_{\rm dir}, f_{c}$ are 
identical. Finally a log-likelihood fit is performed within a fit range 
of $0.1<m_{ee}<0.3~$GeV/$c^{2}$ to determine the direct virtual photon 
fraction for several \pt bins 
separately [$1<p_T<1.5, 1.5<p_T<2.0, 2.0<p_T<3.0, 3.0<p_T<5.0$~GeV/$c$].

\begin{figure}[thb]
\includegraphics[width=1.0\linewidth]{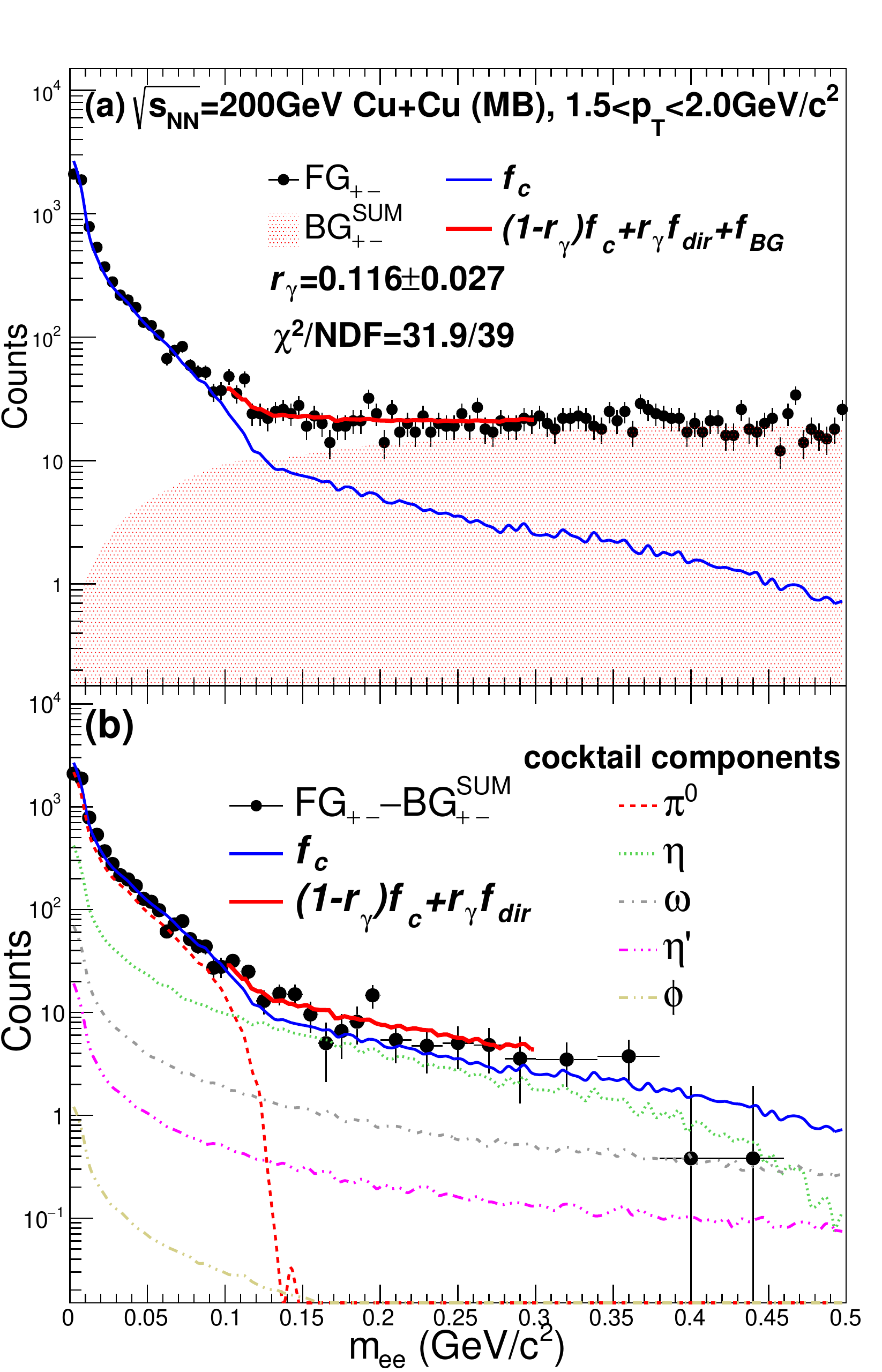} 
\caption{\label{fig:det_r_gamma} 
The $\ee$ pair mass distribution in Cu$+$Cu MB collisions for 
$1.5<\pt<2.0$~GeV$/c$.  (a) The data (closed [black] circles), fit to 
the data $(1-r_{\gamma})f_{c}+r_{\gamma}f_{\rm dir}+f_{\rm BG}$ (thick 
[red] curve), hadronic contribution (thin [blue] curve), and BG$^{\rm 
SUM}_{+-}$ (shaded [red] region). (b) The data after BG$^{\rm SUM}_{+-}$ 
subtraction (closed [black] circles), the fit (thick [red] curve), 
hadronic contribution (thin [blue] curve), and cocktail 
components (indicated curves [see legend]).}
\end{figure}

Figure~\ref{fig:det_r_gamma} shows the $\ee$ pair mass distributions in 
Cu$+$Cu MB collisions for $1.5<\pt<2.0$~GeV$/c$.  
Figure~\ref{fig:det_r_gamma}a shows the data, the fit, the hadronic 
contribution, and the background BG$^{\rm SUM}_{+-}$.  
Figure~\ref{fig:det_r_gamma}b shows the data and fit after 
BG$^{\rm SUM}_{+-}$ subtraction, the hadronic contribution, and cocktail 
components.

\subsection{Systematic uncertainties}

The major sources of systematic uncertainties of the direct photon 
fraction are:

\begin{enumerate}
\item the background normalization,
\item the particle composition of the hadronic cocktail,
\item the $\ee$ mass range for the log-likelihood fit.
\end{enumerate}

\noindent To evaluate the uncertainty of the direct photon fraction, the 
fraction is recalculated by the same procedure varying each source 
within $\pm1\sigma$ of its uncertainty. The differences from the nominal 
value are quantified and taken as contributions to the uncertainty of 
the direct photon fraction. An uncertainty of about 15\%--40\% comes from 
the fit mass range with different fit starting points from 0 to 
0.15~GeV/$c^2$. The particle compositions, dominantly $\pi^{0}/\eta$ add 
another 5\%--15\%. An additional 9.6\% and 10\% uncertainties are introduced 
from the MB trigger efficiency and $\ee$ pair acceptance when converting 
the direct photon fraction to the yield. Total systematic uncertainties 
are calculated as a quadratic sum.

\begin{figure*}[thb]
\includegraphics[width=0.99\linewidth]{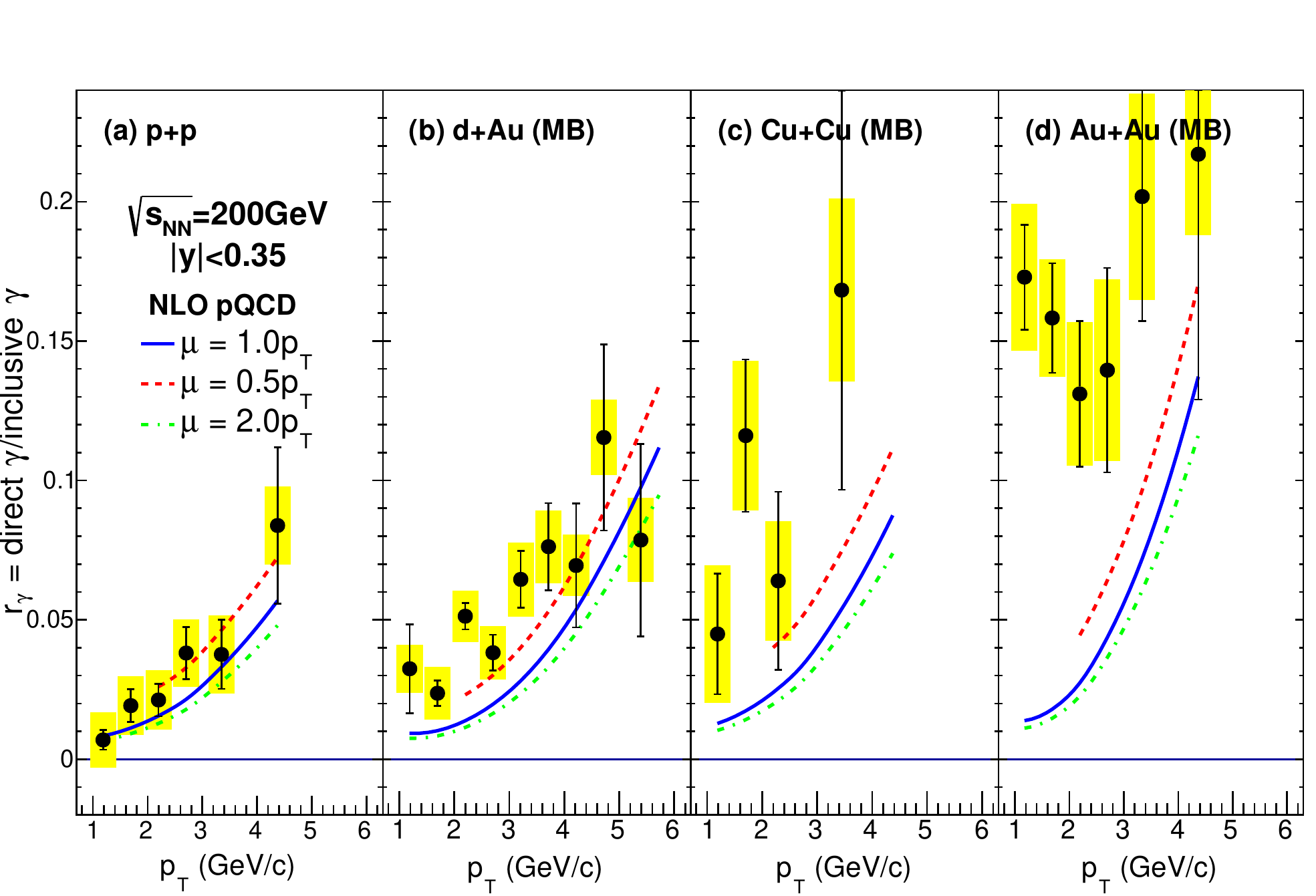}
\caption{\label{fig:comp_r_gamma}
Direct photon fraction measured with the virtual photon method for 
different systems in $\sqsn=200~$GeV collisions: (a) 
$p$$+$$p$~\cite{Adare:2012vn}, (b) $d$$+$Au (MB)~\cite{Adare:2012vn}, (c) 
Cu$+$Cu (MB), (d) Au$+$Au (MB)~\cite{Adare:2008ab}. Expectations from NLO 
pQCD calculations~\cite{Gordon:1993qc} are also shown by curves with 
different cutoff mass scales, $\mu$.}
\end{figure*}

\section{Results and Discussion}

\begin{figure}[thb]
\includegraphics[width=1.0\linewidth]{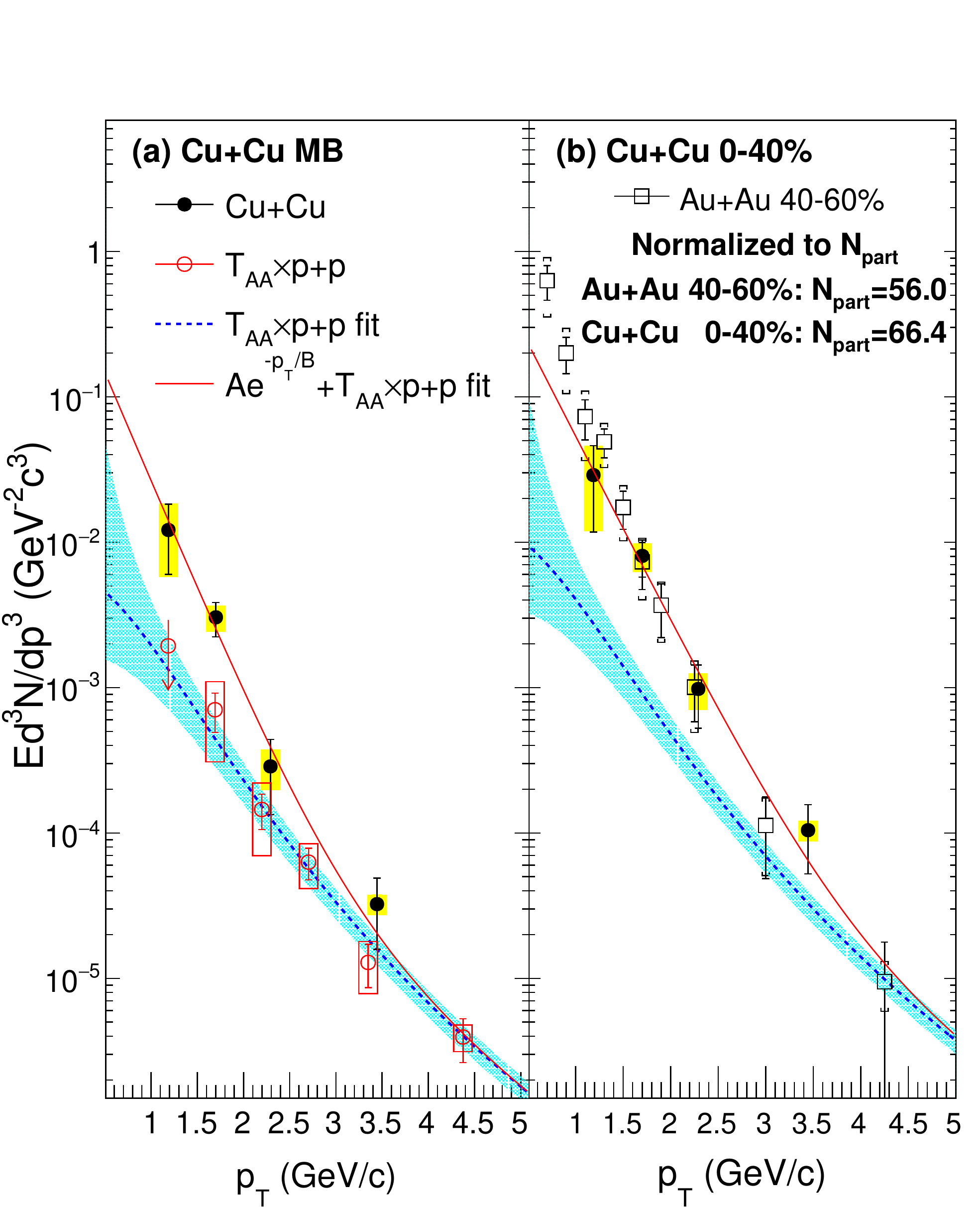}
\caption{\label{fig:y_gamma}
The direct photon spectra (closed [black] circles) for 200~GeV Cu$+$Cu 
(a) MB and (b) 0\%--40\% centralities. The $\taa$-scaled $p$$+$$p$ data 
and fits together with uncertainties are shown as the open [red] circles 
symbols and the dotted [blue] curves and accompanying [red] boxes and 
[blue] bands. Au$+$Au 40\%--60\% centrality data points, which have a 
similar $\Npart$ as the Cu$+$Cu 0\%-40\% centrality data, are shown as 
the open [black] squares, where the Au$+$Au points are scaled by 
the $\Npart$ ratio (66.4/56.0).  An exponential fit to the Cu$+$Cu data 
of the excess yield over the scaled $p$$+$$p$ fit (solid [red] curve) 
yields inverse slopes of 285$\pm$53(stat)$\pm$57(syst)~MeV/$c$ for MB 
and 333$\pm$72(stat)$\pm$45(syst)~MeV/$c$ for 0\%--40\%.
}
\end{figure}

\begin{table*}[tbh]
\caption{$dN_{\rm ch}/d\eta$, $\Ncoll$, $\Npart$, the inverse slope of the 
exponential fits, and $dN/dy(\pt>1~$GeV$/c)$ of the excess yield of 
direct photons over the scaled $p$$+$$p$ fits for 0\%--40\% and MB Cu$+$Cu 
collisions.}
\begin{ruledtabular} \begin{tabular}{lccccc}
Centrality & $dN_{\rm ch}/d\eta$ & $\Ncoll$       & $\Npart$     & Inverse slope (MeV/$c$) & $dN/dy(\pt>1~$GeV$/c)$ \\
\hline
0\%--40\%     & 109.3$\pm$7.8   & 108.2$\pm$12.0 & 66.4$\pm$2.5 & 333$\pm$72$\pm$45       & (1.3$\pm$0.5$^{+0.9}_{-0.8})\times10^{-1}$ \\
MB         &  51.7$\pm$3.6   & 51.8$\pm$5.6   & 34.6$\pm$1.2 & 285$\pm$53$\pm$57       & (5.4$\pm$1.9$^{+3.6}_{-3.1})\times10^{-2}$ \\
\end{tabular} \end{ruledtabular}
\label{tb:sum_data}
\end{table*}

\subsection{Direct-photon fraction}

The direct photon fraction as a function of \pt is obtained for two 
different centrality classes, MB and 0\%--40\%. 
Figure~\ref{fig:comp_r_gamma} shows the comparison of the direct photon 
fraction, $r_{\gamma}$, measured with the virtual photon method for 
different collision systems at $\sqsn=200~$GeV from left to right: 
$p$$+$$p$~\cite{Adare:2012vn}, $d$$+$Au (MB)~\cite{Adare:2012vn}, 
Cu$+$Cu (MB), and Au$+$Au (MB)~\cite{Adare:2008ab}.

The statistical and systematic uncertainties are shown together with the 
data points. Curves indicate the expectations from a 
next-to-leading-order (NLO) perturbative-quantum-chromodynamics (pQCD)
calculation~\cite{Gordon:1993qc} with different cutoff mass scales, 
$\mu$. While the $p$$+$$p$ and $d$$+$Au results show agreements with the NLO 
pQCD calculation, an excess over the NLO pQCD calculation is seen in the 
Cu$+$Cu data as well as in Au$+$Au. The Cu$+$Cu excess is rather modest 
compared to Au$+$Au, possibly due to a smaller volume of the created 
medium.

\subsection{Direct photon spectra}

The obtained direct photon fractions are converted to direct photon 
yields using the inclusive photon yields calculated by the same Monte 
Carlo simulation used for the $\ee$ pairs of the hadronic cocktail. 
Figure~\ref{fig:y_gamma} shows the direct photon spectra for Cu$+$Cu MB 
and 0\%--40\% most central events. The $p$$+$$p$ 
results~\cite{Adare:2012vn} parameterized by a modified power law 
function, $A_{pp}(1+\pt^{2}/B_{pp})^{n_{pp}}$, and its $\taa$-scaled 
functions are shown as the dotted curves together with the data points. 
The modified power law is an empirical parameterization describing the 
$p$$+$$p$ result well, especially at low \pt. The same function has been 
employed in previous low \pt direct photon publications in heavy ion 
collisions~\cite{Adare:2008ab,Adare:2014fwh}. We have performed a least 
square analysis in which \pt-correlated and \pt-uncorrelated errors are 
properly taken into account. A detail description on constraint 
parameterization can be found in Ref.~\cite{Adare:2008cg}. The $p$$+$$p$ 
data points measured by the EMCal in $4<\pt<10~$GeV/$c$ are included in 
the fit in addition to the virtual photon measurement covering $\pt <$ 
6~GeV/$c$. Here the lowest \pt data point is just an upper limit. The 
best fit gives ${\chi}^2$/NDF=18.9/17, which is the minimum obtained by 
variation of the \pt-correlated errors. The uncertainty of the $p$$+$$p$ 
fit is calculated using the error matrix of the fit parameters and is 
indicated as bands on the scaled $p$$+$$p$ fits. A different empirical 
parameterization, employed in Ref.~\cite{Adare:2012vn}, was tested as 
well.  We treat the small deviation we find above 1~GeV/$c$ as a 
maximum-extend error. We divide the deviation by $\sqrt{12}$ and add it 
in quadrature to the uncertainty of the fit.

An exponential fit to the excess yield above the scaled $p$$+$$p$ fits 
gives inverse slopes of 285$\pm$53(stat)$\pm$57(syst)~MeV/$c$ for MB and 
333$\pm$72(stat)$\pm$45(syst)~MeV/$c$ for 0\%--40\% centrality. 
Furthermore, the Cu$+$Cu 0\%--40\% centrality result is compared with 
the Au$+$Au 40\%--60\% data scaled by the $\Npart$ ratio (66.4/56.0), 
which is consistent within uncertainties [see 
Fig.~\ref{fig:y_gamma}(b)].

\subsection{Rapidity density}

\begin{figure}[thb]
\includegraphics[width=1.0\linewidth]{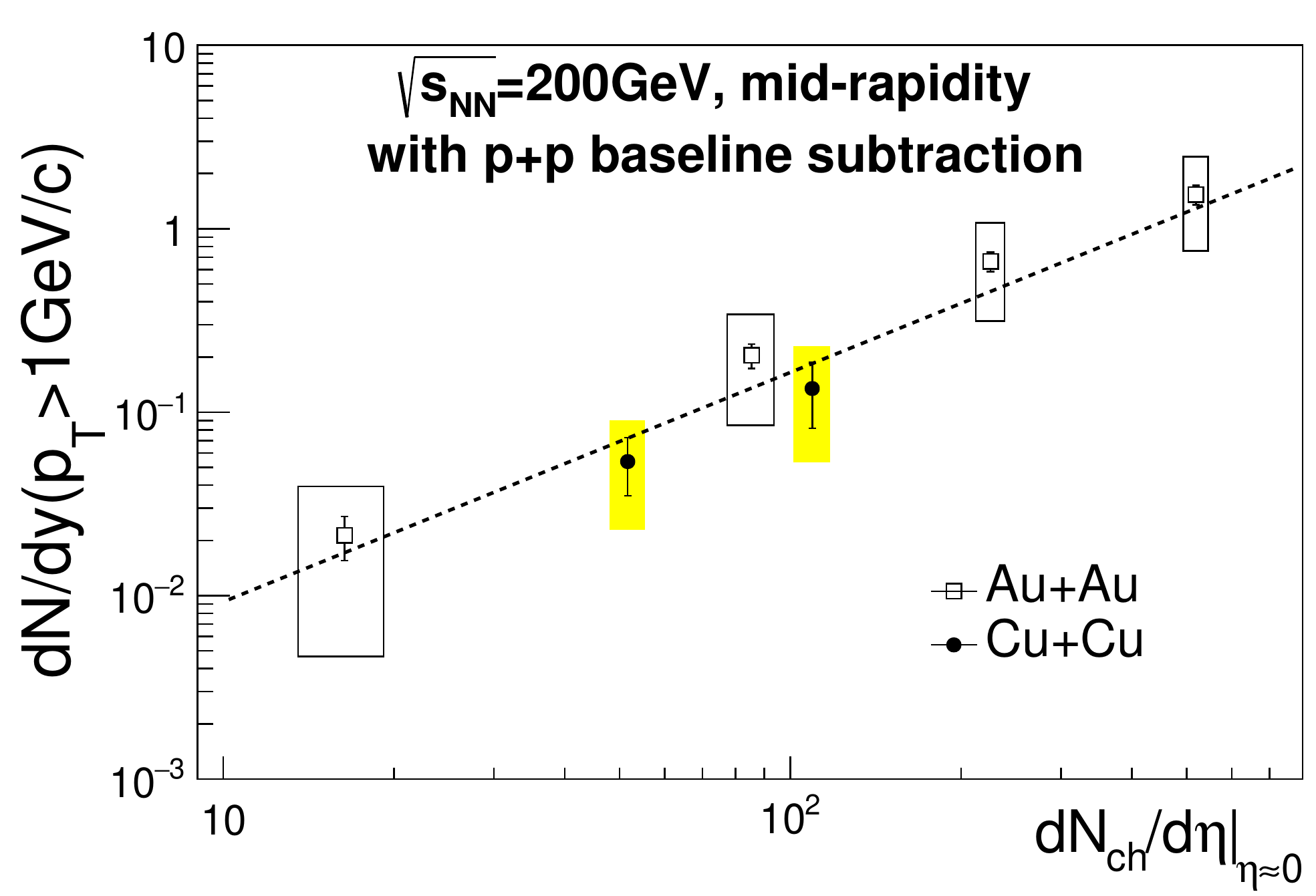}
\caption{\label{fig:comp_dNdy}
Rapidity densities of the excess yield of direct photons over the scaled 
$p$$+$$p$ fits for $p_{T}>1~$GeV/$c$ at midrapidity as a function of 
$dN_{\rm ch}/d\eta$. The Au$+$Au data points with different 
centralities~\cite{Adare:2014fwh} and the power-law fit with the fixed 
power of 1.25 to both Cu$+$Cu and Au$+$Au data points are shown together.}
\end{figure}

We further investigate the $\Npart$ dependence of the direct photon 
yields as discussed in Ref.~\cite{Adare:2014fwh}. It has been reported 
that the Au$+$Au results~\cite{PPG212} show an increasing trend for 
$\Npart$. The Cu$+$Cu data points help to have a closer look at the 
dependence in the small $\Npart$ region. The rapidity density for 
$\pt>1~$GeV/$c$ at midrapidity, $dN/dy(\pt>1~$GeV$/c)$, is calculated 
by summing the direct photon yields in given \pt bins taking the 
bin-width correction into account:

\begin{eqnarray}
\frac{dN}{dy} & = & 2\pi\sum_{\pt^{i}>1\rm{GeV}/c}(\pt^{i}\times y_{\gamma}^{i}\times C_{bw}^{i} \times \Delta\pt^{i}), \\
C_{bw}^{i}    & = & \int_{p_{T,min}}^{p_{T,max}}f_{fit}(\pt)d\pt/(f_{fit}(\pt^{i})\times\Delta\pt^{i}),
\label{eq:bw_cor}
\end{eqnarray}

\noindent where $\pt^i, y_{\gamma}^i, \Delta\pt^{i}$ are the mean \pt, 
the direct photon yield and the \pt-bin width for the $i$-th \pt 
bin. The bin-width correction, $C_{bw}$, is evaluated based on the fit 
function, $f_{fit}$, to the data shown in Fig.~\ref{fig:y_gamma}. 
$C_{bw}$ contributes an additional 3.5\% uncertainty of $dN/dy$. Then, 
$dN/dy$ for the binary-scaled $p$$+$$p$ fit~\cite{PPG212} is subtracted. 
Figure~\ref{fig:comp_dNdy} shows $dN/dy$ of the excess yield over the 
scaled $p$$+$$p$ fit as a function of measured charged multiplicity, 
$dN_{\rm ch}/d\eta$, at midrapidity. A simple power law fit with the fixed 
power of 1.25, $(dN_{\rm ch}/d\eta)^{1.25}$, is done for both the Cu$+$Cu and 
Au$+$Au results as done in Ref.~\cite{PPG212}. It works very well to 
describe the $dN_{\rm ch}/d\eta$ dependence.

The inverse slope of the exponential fits and the rapidity density of 
the excess yield of direct photons over the scaled $p$$+$$p$ fits for 
$p_{T}>1~$GeV/$c$ are summarized together with $dN_{\rm ch}/d\eta$, 
$\Ncoll$, $\Npart$ corresponding to 0\%--40\%, MB Cu$+$Cu collisions in 
Table~\ref{tb:sum_data}.

\section{Summary and Conclusions}

Low-\pt direct photons have been measured using the virtual photon 
method for MB and 0\%--40\% most central collisions in $\sqsn=200$~GeV 
Cu$+$Cu collisions. A clear excess yield of direct photons over the 
binary-scaled $p$$+$$p$ baseline is seen for Cu$+$Cu as in the previously 
reported Au$+$Au results. The Cu$+$Cu direct photon $p_{T}$ spectra are 
consistent with the Au$+$Au data for similar $\Npart$. The exponential 
fits to the excess over the binary-scaled $p$$+$$p$ baseline give inverse 
slopes of 285$\pm$53(stat)$\pm$57(syst)~MeV/$c$ for MB and 
333$\pm$72(stat)$\pm$45(syst)~MeV/$c$ for 0\%--40\% centrality. 
The Cu$+$Cu data points improve our knowledge of the system size 
dependence of the excess yield of the direct photons, especially in the 
small-$\Npart$ region. The Cu$+$Cu results on $dN/dy$ for $\pt>1~$GeV/$c$ 
follow the same $dN_{\rm ch}/d\eta$ dependence as the Au$+$Au data as 
described by a simple power law.

\begin{acknowledgments}


We thank the staff of the Collider-Accelerator and Physics
Departments at Brookhaven National Laboratory and the staff of
the other PHENIX participating institutions for their vital
contributions.  We acknowledge support from the 
Office of Nuclear Physics in the
Office of Science of the Department of Energy,
the National Science Foundation, 
Abilene Christian University Research Council, 
Research Foundation of SUNY, and
Dean of the College of Arts and Sciences, Vanderbilt University 
(U.S.A),
Ministry of Education, Culture, Sports, Science, and Technology
and the Japan Society for the Promotion of Science (Japan),
Conselho Nacional de Desenvolvimento Cient\'{\i}fico e
Tecnol{\'o}gico and Funda\c c{\~a}o de Amparo {\`a} Pesquisa do
Estado de S{\~a}o Paulo (Brazil),
Natural Science Foundation of China (People's Republic of China),
Croatian Science Foundation and
Ministry of Science and Education (Croatia),
Ministry of Education, Youth and Sports (Czech Republic),
Centre National de la Recherche Scientifique, Commissariat
{\`a} l'{\'E}nergie Atomique, and Institut National de Physique
Nucl{\'e}aire et de Physique des Particules (France),
Bundesministerium f\"ur Bildung und Forschung, Deutscher Akademischer 
Austausch Dienst, and Alexander von Humboldt Stiftung (Germany),
J. Bolyai Research Scholarship, EFOP, the New National Excellence
Program ({\'U}NKP), NKFIH, and OTKA (Hungary),
Department of Atomic Energy and Department of Science and Technology 
(India),
Israel Science Foundation (Israel), 
Basic Science Research Program through NRF of the Ministry of 
Education (Korea),
Physics Department, Lahore University of Management Sciences (Pakistan),
Ministry of Education and Science, Russian Academy of Sciences,
Federal Agency of Atomic Energy (Russia),
VR and Wallenberg Foundation (Sweden), 
the U.S. Civilian Research and Development Foundation for the
Independent States of the Former Soviet Union, 
the Hungarian American Enterprise Scholarship Fund,
the US-Hungarian Fulbright Foundation,
and the US-Israel Binational Science Foundation.

\end{acknowledgments}



\begin{thebibliography}{26}%
\makeatletter
\providecommand \@ifxundefined [1]{%
 \@ifx{#1\undefined}
}%
\providecommand \@ifnum [1]{%
 \ifnum #1\expandafter \@firstoftwo
 \else \expandafter \@secondoftwo
 \fi
}%
\providecommand \@ifx [1]{%
 \ifx #1\expandafter \@firstoftwo
 \else \expandafter \@secondoftwo
 \fi
}%
\providecommand \natexlab [1]{#1}%
\providecommand \enquote  [1]{``#1''}%
\providecommand \bibnamefont  [1]{#1}%
\providecommand \bibfnamefont [1]{#1}%
\providecommand \citenamefont [1]{#1}%
\providecommand \href@noop [0]{\@secondoftwo}%
\providecommand \href [0]{\begingroup \@sanitize@url \@href}%
\providecommand \@href[1]{\@@startlink{#1}\@@href}%
\providecommand \@@href[1]{\endgroup#1\@@endlink}%
\providecommand \@sanitize@url [0]{\catcode `\\12\catcode `\$12\catcode
  `\&12\catcode `\#12\catcode `\^12\catcode `\_12\catcode `\%12\relax}%
\providecommand \@@startlink[1]{}%
\providecommand \@@endlink[0]{}%
\providecommand \url  [0]{\begingroup\@sanitize@url \@url }%
\providecommand \@url [1]{\endgroup\@href {#1}{\urlprefix }}%
\providecommand \urlprefix  [0]{URL }%
\providecommand \Eprint [0]{\href }%
\providecommand \doibase [0]{http://dx.doi.org/}%
\providecommand \selectlanguage [0]{\@gobble}%
\providecommand \bibinfo  [0]{\@secondoftwo}%
\providecommand \bibfield  [0]{\@secondoftwo}%
\providecommand \translation [1]{[#1]}%
\providecommand \BibitemOpen [0]{}%
\providecommand \bibitemStop [0]{}%
\providecommand \bibitemNoStop [0]{.\EOS\space}%
\providecommand \EOS [0]{\spacefactor3000\relax}%
\providecommand \BibitemShut  [1]{\csname bibitem#1\endcsname}%
\let\auto@bib@innerbib\@empty
\bibitem [{\citenamefont {Adcox}\ \emph {et~al.}(2005)\citenamefont {Adcox}
  \emph {et~al.}}]{Adcox:2004mh}%
  \BibitemOpen
  \bibfield  {author} {\bibinfo {author} {\bibfnamefont {K.}~\bibnamefont
  {Adcox}} \emph {et~al.} (\bibinfo {collaboration} {PHENIX Collaboration}),\
  }\bibfield  {title} {\enquote {\bibinfo {title} {{Formation of dense partonic
  matter in relativistic nucleus-nucleus collisions at RHIC: Experimental
  evaluation by the PHENIX Collaboration}},}\ }\href {\doibase
  10.1016/j.nuclphysa.2005.03.086} {\bibfield  {journal} {\bibinfo  {journal}
  {Nucl. Phys. A}\ }\textbf {\bibinfo {volume} {757}},\ \bibinfo {pages} {184}
  (\bibinfo {year} {2005})}\BibitemShut {NoStop}%
\bibitem [{\citenamefont {Adams}\ \emph {et~al.}(2005)\citenamefont {Adams}
  \emph {et~al.}}]{Adams:2005dq}%
  \BibitemOpen
  \bibfield  {author} {\bibinfo {author} {\bibfnamefont {John}\ \bibnamefont
  {Adams}} \emph {et~al.} (\bibinfo {collaboration} {STAR Collaboration}),\
  }\bibfield  {title} {\enquote {\bibinfo {title} {{Experimental and
  theoretical challenges in the search for the quark gluon plasma: The STAR
  Collaboration's critical assessment of the evidence from RHIC collisions}},}\
  }\href {\doibase 10.1016/j.nuclphysa.2005.03.085} {\bibfield  {journal}
  {\bibinfo  {journal} {Nucl. Phys. A}\ }\textbf {\bibinfo {volume} {757}},\
  \bibinfo {pages} {102} (\bibinfo {year} {2005})}\BibitemShut {NoStop}%
\bibitem [{\citenamefont {David}\ \emph {et~al.}(2008)\citenamefont {David},
  \citenamefont {Rapp},\ and\ \citenamefont {Xu}}]{David:2006sr}%
  \BibitemOpen
  \bibfield  {author} {\bibinfo {author} {\bibfnamefont {G.}~\bibnamefont
  {David}}, \bibinfo {author} {\bibfnamefont {R.}~\bibnamefont {Rapp}}, \ and\
  \bibinfo {author} {\bibfnamefont {Z.}~\bibnamefont {Xu}},\ }\bibfield
  {title} {\enquote {\bibinfo {title} {{Electromagnetic Probes at RHIC-II}},}\
  }\href {\doibase 10.1016/j.physrep.2008.04.003} {\bibfield  {journal}
  {\bibinfo  {journal} {Phys. Rept.}\ }\textbf {\bibinfo {volume} {462}},\
  \bibinfo {pages} {176} (\bibinfo {year} {2008})}\BibitemShut {NoStop}%
\bibitem [{\citenamefont {Stankus}(2005)}]{Stankus:2005eq}%
  \BibitemOpen
  \bibfield  {author} {\bibinfo {author} {\bibfnamefont {P.}~\bibnamefont
  {Stankus}},\ }\bibfield  {title} {\enquote {\bibinfo {title} {{Direct photon
  production in relativistic heavy-ion collisions}},}\ }\href {\doibase
  10.1146/annurev.nucl.53.041002.110533} {\bibfield  {journal} {\bibinfo
  {journal} {Ann. Rev. Nucl. Part. Sci.}\ }\textbf {\bibinfo {volume} {55}},\
  \bibinfo {pages} {517} (\bibinfo {year} {2005})}\BibitemShut {NoStop}%
\bibitem [{\citenamefont {Turbide}\ \emph {et~al.}(2004)\citenamefont
  {Turbide}, \citenamefont {Rapp},\ and\ \citenamefont
  {Gale}}]{Turbide:2003si}%
  \BibitemOpen
  \bibfield  {author} {\bibinfo {author} {\bibfnamefont {S.}~\bibnamefont
  {Turbide}}, \bibinfo {author} {\bibfnamefont {R.}~\bibnamefont {Rapp}}, \
  and\ \bibinfo {author} {\bibfnamefont {C.}~\bibnamefont {Gale}},\ }\bibfield
  {title} {\enquote {\bibinfo {title} {{Hadronic production of thermal
  photons}},}\ }\href {\doibase 10.1103/PhysRevC.69.014903} {\bibfield
  {journal} {\bibinfo  {journal} {Phys. Rev. C}\ }\textbf {\bibinfo {volume}
  {69}},\ \bibinfo {pages} {014903} (\bibinfo {year} {2004})}\BibitemShut
  {NoStop}%
\bibitem [{\citenamefont {Adare}\ \emph
  {et~al.}(2010{\natexlab{a}})\citenamefont {Adare} \emph
  {et~al.}}]{Adare:2008ab}%
  \BibitemOpen
  \bibfield  {author} {\bibinfo {author} {\bibfnamefont {A.}~\bibnamefont
  {Adare}} \emph {et~al.} (\bibinfo {collaboration} {PHENIX Collaboration}),\
  }\bibfield  {title} {\enquote {\bibinfo {title} {{Enhanced production of
  direct photons in Au+Au collisions at $\sqrt{s_{NN}}=200$ GeV and
  implications for the initial temperature}},}\ }\href {\doibase
  10.1103/PhysRevLett.104.132301} {\bibfield  {journal} {\bibinfo  {journal}
  {Phys. Rev. Lett.}\ }\textbf {\bibinfo {volume} {104}},\ \bibinfo {pages}
  {132301} (\bibinfo {year} {2010}{\natexlab{a}})}\BibitemShut {NoStop}%
\bibitem [{\citenamefont {Adare}\ \emph
  {et~al.}(2015{\natexlab{a}})\citenamefont {Adare} \emph
  {et~al.}}]{Adare:2014fwh}%
  \BibitemOpen
  \bibfield  {author} {\bibinfo {author} {\bibfnamefont {A.}~\bibnamefont
  {Adare}} \emph {et~al.} (\bibinfo {collaboration} {PHENIX Collaboration}),\
  }\bibfield  {title} {\enquote {\bibinfo {title} {{Centrality dependence of
  low-momentum direct-photon production in Au$+$Au collisions at
  $\sqrt{s_{_{NN}}}=200$ GeV}},}\ }\href {\doibase 10.1103/PhysRevC.91.064904}
  {\bibfield  {journal} {\bibinfo  {journal} {Phys. Rev. C}\ }\textbf {\bibinfo
  {volume} {91}},\ \bibinfo {pages} {064904} (\bibinfo {year}
  {2015}{\natexlab{a}})}\BibitemShut {NoStop}%
\bibitem [{\citenamefont {Adamczyk}\ \emph {et~al.}(2017)\citenamefont
  {Adamczyk} \emph {et~al.}}]{STAR:2016use}%
  \BibitemOpen
  \bibfield  {author} {\bibinfo {author} {\bibfnamefont {L.}~\bibnamefont
  {Adamczyk}} \emph {et~al.} (\bibinfo {collaboration} {STAR Collaboration}),\
  }\bibfield  {title} {\enquote {\bibinfo {title} {{Direct virtual photon
  production in Au+Au collisions at $\sqrt{s_{NN}}=200$ GeV}},}\ }\href
  {\doibase 10.1016/j.physletb.2017.04.050} {\bibfield  {journal} {\bibinfo
  {journal} {Phys. Lett. B}\ }\textbf {\bibinfo {volume} {770}},\ \bibinfo
  {pages} {451} (\bibinfo {year} {2017})}\BibitemShut {NoStop}%
\bibitem [{\citenamefont {Adare}\ \emph {et~al.}(2013)\citenamefont {Adare}
  \emph {et~al.}}]{Adare:2012vn}%
  \BibitemOpen
  \bibfield  {author} {\bibinfo {author} {\bibfnamefont {A.}~\bibnamefont
  {Adare}} \emph {et~al.} (\bibinfo {collaboration} {PHENIX Collaboration}),\
  }\bibfield  {title} {\enquote {\bibinfo {title} {{Direct photon production in
  $d+$Au collisions at $\sqrt{s_{NN}}=200$ GeV}},}\ }\href {\doibase
  10.1103/PhysRevC.87.054907} {\bibfield  {journal} {\bibinfo  {journal} {Phys.
  Rev. C}\ }\textbf {\bibinfo {volume} {87}},\ \bibinfo {pages} {054907}
  (\bibinfo {year} {2013})}\BibitemShut {NoStop}%
\bibitem [{\citenamefont {Adam}\ \emph {et~al.}(2016)\citenamefont {Adam} \emph
  {et~al.}}]{Adam:2015lda}%
  \BibitemOpen
  \bibfield  {author} {\bibinfo {author} {\bibfnamefont {J.}~\bibnamefont
  {Adam}} \emph {et~al.} (\bibinfo {collaboration} {ALICE Collaboration}),\
  }\bibfield  {title} {\enquote {\bibinfo {title} {{Direct photon production in
  Pb$-$Pb collisions at $\sqrt{s_{NN}}=2.76$ TeV}},}\ }\href {\doibase
  10.1016/j.physletb.2016.01.020} {\bibfield  {journal} {\bibinfo  {journal}
  {Phys. Lett. B}\ }\textbf {\bibinfo {volume} {754}},\ \bibinfo {pages} {235}
  (\bibinfo {year} {2016})}\BibitemShut {NoStop}%
\bibitem [{\citenamefont {Adare}\ \emph {et~al.}(2014)\citenamefont {Adare}
  \emph {et~al.}}]{Adare:2013yxp}%
  \BibitemOpen
  \bibfield  {author} {\bibinfo {author} {\bibfnamefont {A.}~\bibnamefont
  {Adare}} \emph {et~al.} (\bibinfo {collaboration} {PHENIX Collaboration}),\
  }\bibfield  {title} {\enquote {\bibinfo {title} {{System-size dependence of
  open-heavy-flavor production in nucleus-nucleus collisions at
  $\sqrt{s_{_{NN}}}=200$ GeV}},}\ }\href {\doibase 10.1103/PhysRevC.90.034903}
  {\bibfield  {journal} {\bibinfo  {journal} {Phys. Rev. C}\ }\textbf {\bibinfo
  {volume} {90}},\ \bibinfo {pages} {034903} (\bibinfo {year}
  {2014})}\BibitemShut {NoStop}%
\bibitem [{\citenamefont {Adare}\ \emph
  {et~al.}(2016{\natexlab{a}})\citenamefont {Adare} \emph
  {et~al.}}]{Adare:2015bua}%
  \BibitemOpen
  \bibfield  {author} {\bibinfo {author} {\bibfnamefont {A.}~\bibnamefont
  {Adare}} \emph {et~al.} (\bibinfo {collaboration} {PHENIX Collaboration}),\
  }\bibfield  {title} {\enquote {\bibinfo {title} {{Transverse energy
  production and charged-particle multiplicity at midrapidity in various
  systems from $\sqrt{s_{NN}}=7.7$ to 200 GeV}},}\ }\href {\doibase
  10.1103/PhysRevC.93.024901} {\bibfield  {journal} {\bibinfo  {journal} {Phys.
  Rev. C}\ }\textbf {\bibinfo {volume} {93}},\ \bibinfo {pages} {024901}
  (\bibinfo {year} {2016}{\natexlab{a}})}\BibitemShut {NoStop}%
\bibitem [{\citenamefont {Adcox}\ \emph {et~al.}(2003)\citenamefont {Adcox}
  \emph {et~al.}}]{PHENIX-NIM}%
  \BibitemOpen
  \bibfield  {author} {\bibinfo {author} {\bibfnamefont {K.}~\bibnamefont
  {Adcox}} \emph {et~al.} (\bibinfo {collaboration} {PHENIX Collaboration}),\
  }\bibfield  {title} {\enquote {\bibinfo {title} {{PHENIX detector
  overview}},}\ }\href {\doibase 10.1016/S0168-9002(02)01950-2} {\bibfield
  {journal} {\bibinfo  {journal} {Nucl. Instrum. Methods Phys. Res., Sec. A}\
  }\textbf {\bibinfo {volume} {499}},\ \bibinfo {pages} {469} (\bibinfo {year}
  {2003})}\BibitemShut {NoStop}%
\bibitem [{\citenamefont {Adare}\ \emph
  {et~al.}(2015{\natexlab{b}})\citenamefont {Adare} \emph
  {et~al.}}]{Adare:2014mgk}%
  \BibitemOpen
  \bibfield  {author} {\bibinfo {author} {\bibfnamefont {A.}~\bibnamefont
  {Adare}} \emph {et~al.} (\bibinfo {collaboration} {PHENIX Collaboration}),\
  }\bibfield  {title} {\enquote {\bibinfo {title} {{Search for dark photons
  from neutral meson decays in $p$$+$$p$ and $d$$+$Au collisions at
  $\sqrt{s_{NN}}=200$ GeV}},}\ }\href {\doibase 10.1103/PhysRevC.91.031901}
  {\bibfield  {journal} {\bibinfo  {journal} {Phys. Rev. C}\ }\textbf {\bibinfo
  {volume} {91}},\ \bibinfo {pages} {031901} (\bibinfo {year}
  {2015}{\natexlab{b}})}\BibitemShut {NoStop}%
\bibitem [{\citenamefont {Adare}\ \emph
  {et~al.}(2010{\natexlab{b}})\citenamefont {Adare} \emph
  {et~al.}}]{Adare:2009qk}%
  \BibitemOpen
  \bibfield  {author} {\bibinfo {author} {\bibfnamefont {A.}~\bibnamefont
  {Adare}} \emph {et~al.} (\bibinfo {collaboration} {PHENIX Collaboration}),\
  }\bibfield  {title} {\enquote {\bibinfo {title} {{Detailed measurement of the
  $e^{+}e^{-}$ pair continuum in $p+p$ and Au+Au collisions at
  $\sqrt{s_{NN}}=200$ GeV and implications for direct photon production}},}\
  }\href {\doibase 10.1103/PhysRevC.81.034911} {\bibfield  {journal} {\bibinfo
  {journal} {Phys. Rev. C}\ }\textbf {\bibinfo {volume} {81}},\ \bibinfo
  {pages} {034911} (\bibinfo {year} {2010}{\natexlab{b}})}\BibitemShut
  {NoStop}%
\bibitem [{\citenamefont {Adare}\ \emph
  {et~al.}(2016{\natexlab{b}})\citenamefont {Adare} \emph
  {et~al.}}]{Adare:2015ila}%
  \BibitemOpen
  \bibfield  {author} {\bibinfo {author} {\bibfnamefont {A.}~\bibnamefont
  {Adare}} \emph {et~al.} (\bibinfo {collaboration} {PHENIX Collaboration}),\
  }\bibfield  {title} {\enquote {\bibinfo {title} {{Dielectron production in
  Au$+$Au collisions at $\sqrt{s_{NN}}=200$ GeV}},}\ }\href {\doibase
  10.1103/PhysRevC.93.014904} {\bibfield  {journal} {\bibinfo  {journal} {Phys.
  Rev. C}\ }\textbf {\bibinfo {volume} {93}},\ \bibinfo {pages} {014904}
  (\bibinfo {year} {2016}{\natexlab{b}})}\BibitemShut {NoStop}%
\bibitem [{\citenamefont {Sjostrand}\ \emph {et~al.}()\citenamefont
  {Sjostrand}, \citenamefont {Mrenna},\ and\ \citenamefont
  {Skands}}]{Sjostrand:2006za}%
  \BibitemOpen
  \bibfield  {author} {\bibinfo {author} {\bibfnamefont {T.}~\bibnamefont
  {Sjostrand}}, \bibinfo {author} {\bibfnamefont {S.}~\bibnamefont {Mrenna}}, \
  and\ \bibinfo {author} {\bibfnamefont {P.~Z.}\ \bibnamefont {Skands}},\
  }\href@noop {} {\enquote {\bibinfo {title} {{PYTHIA 6.4 Physics and
  Manual}},}\ }\bibinfo {note} {{J. High Energy Phys. {\bf 05 (2006)}
  026}}\BibitemShut {NoStop}%
\bibitem [{\citenamefont {Sjostrand}\ \emph {et~al.}(2008)\citenamefont
  {Sjostrand}, \citenamefont {Mrenna},\ and\ \citenamefont
  {Skands}}]{Sjostrand:2007gs}%
  \BibitemOpen
  \bibfield  {author} {\bibinfo {author} {\bibfnamefont {T.}~\bibnamefont
  {Sjostrand}}, \bibinfo {author} {\bibfnamefont {S.}~\bibnamefont {Mrenna}}, \
  and\ \bibinfo {author} {\bibfnamefont {P.~Z.}\ \bibnamefont {Skands}},\
  }\bibfield  {title} {\enquote {\bibinfo {title} {{A Brief Introduction to
  PYTHIA 8.1}},}\ }\href {\doibase 10.1016/j.cpc.2008.01.036} {\bibfield
  {journal} {\bibinfo  {journal} {Comput. Phys. Commun.}\ }\textbf {\bibinfo
  {volume} {178}},\ \bibinfo {pages} {852} (\bibinfo {year}
  {2008})}\BibitemShut {NoStop}%
\bibitem [{\citenamefont {Lai}\ \emph {et~al.}(2000)\citenamefont {Lai},
  \citenamefont {Huston}, \citenamefont {Kuhlmann}, \citenamefont {Morfin},
  \citenamefont {Olness}, \citenamefont {Owens}, \citenamefont {Pumplin},\ and\
  \citenamefont {Tung}}]{Lai:1999wy}%
  \BibitemOpen
  \bibfield  {author} {\bibinfo {author} {\bibfnamefont {H.~L.}\ \bibnamefont
  {Lai}}, \bibinfo {author} {\bibfnamefont {J.}~\bibnamefont {Huston}},
  \bibinfo {author} {\bibfnamefont {S.}~\bibnamefont {Kuhlmann}}, \bibinfo
  {author} {\bibfnamefont {J.}~\bibnamefont {Morfin}}, \bibinfo {author}
  {\bibfnamefont {Fredrick~I.}\ \bibnamefont {Olness}}, \bibinfo {author}
  {\bibfnamefont {J.~F.}\ \bibnamefont {Owens}}, \bibinfo {author}
  {\bibfnamefont {J.}~\bibnamefont {Pumplin}}, \ and\ \bibinfo {author}
  {\bibfnamefont {W.~K.}\ \bibnamefont {Tung}} (\bibinfo {collaboration} {CTEQ
  Collaboration}),\ }\bibfield  {title} {\enquote {\bibinfo {title} {{Global
  QCD analysis of parton structure of the nucleon: CTEQ5 parton
  distributions}},}\ }\href {\doibase 10.1007/s100529900196} {\bibfield
  {journal} {\bibinfo  {journal} {Eur. Phys. J. C}\ }\textbf {\bibinfo {volume}
  {12}},\ \bibinfo {pages} {375} (\bibinfo {year} {2000})}\BibitemShut
  {NoStop}%
\bibitem [{\citenamefont {Brun}\ \emph {et~al.}(1987)\citenamefont {Brun},
  \citenamefont {Bruyant}, \citenamefont {Maire}, \citenamefont {McPherson},\
  and\ \citenamefont {Zanarini}}]{Brun:1987ma}%
  \BibitemOpen
  \bibfield  {author} {\bibinfo {author} {\bibfnamefont {R.}~\bibnamefont
  {Brun}}, \bibinfo {author} {\bibfnamefont {F.}~\bibnamefont {Bruyant}},
  \bibinfo {author} {\bibfnamefont {M.}~\bibnamefont {Maire}}, \bibinfo
  {author} {\bibfnamefont {A.~C.}\ \bibnamefont {McPherson}}, \ and\ \bibinfo
  {author} {\bibfnamefont {P.}~\bibnamefont {Zanarini}},\ }\href@noop {}
  {\enquote {\bibinfo {title} {{{\sc geant3}}},}\ } (\bibinfo {year}
  {1987})\BibitemShut {NoStop}%
\bibitem [{\citenamefont {Adare}\ \emph {et~al.}(2017)\citenamefont {Adare}
  \emph {et~al.}}]{Adare:2017caq}%
  \BibitemOpen
  \bibfield  {author} {\bibinfo {author} {\bibfnamefont {A.}~\bibnamefont
  {Adare}} \emph {et~al.} (\bibinfo {collaboration} {PHENIX Collaboration}),\
  }\bibfield  {title} {\enquote {\bibinfo {title} {{Measurements of $e^+e^-$
  pairs from open heavy flavor in $p$$+$$p$ and $d$$+$$A$ collisions at
  $\sqrt{s_{NN}}=200$ GeV}},}\ }\href {\doibase 10.1103/PhysRevC.96.024907}
  {\bibfield  {journal} {\bibinfo  {journal} {Phys. Rev. C}\ }\textbf {\bibinfo
  {volume} {96}},\ \bibinfo {pages} {024907} (\bibinfo {year}
  {2017})}\BibitemShut {NoStop}%
\bibitem [{\citenamefont {Adare}\ \emph
  {et~al.}(2008{\natexlab{a}})\citenamefont {Adare} \emph
  {et~al.}}]{Adare:2008ad}%
  \BibitemOpen
  \bibfield  {author} {\bibinfo {author} {\bibfnamefont {A.}~\bibnamefont
  {Adare}} \emph {et~al.} (\bibinfo {collaboration} {PHENIX Collaboration}),\
  }\bibfield  {title} {\enquote {\bibinfo {title} {{Onset of pi0 Suppression
  Studied in Cu+Cu Collisions at $\sqrt{s_{NN}}=22.4$, 62.4, and 200 GeV}},}\
  }\href {\doibase 10.1103/PhysRevLett.101.162301} {\bibfield  {journal}
  {\bibinfo  {journal} {Phys. Rev. Lett.}\ }\textbf {\bibinfo {volume} {101}},\
  \bibinfo {pages} {162301} (\bibinfo {year} {2008}{\natexlab{a}})}\BibitemShut
  {NoStop}%
\bibitem [{\citenamefont {Adler}\ \emph {et~al.}(2006)\citenamefont {Adler}
  \emph {et~al.}}]{Adler:2006hu}%
  \BibitemOpen
  \bibfield  {author} {\bibinfo {author} {\bibfnamefont {S.~S.}\ \bibnamefont
  {Adler}} \emph {et~al.} (\bibinfo {collaboration} {PHENIX Collaboration}),\
  }\bibfield  {title} {\enquote {\bibinfo {title} {{Common suppression pattern
  of eta and pi0 mesons at high transverse momentum in Au+Au collisions at
  $\sqrt{s_{NN}}=200$ GeV}},}\ }\href {\doibase 10.1103/PhysRevLett.96.202301}
  {\bibfield  {journal} {\bibinfo  {journal} {Phys. Rev. Lett.}\ }\textbf
  {\bibinfo {volume} {96}},\ \bibinfo {pages} {202301} (\bibinfo {year}
  {2006})}\BibitemShut {NoStop}%
\bibitem [{\citenamefont {Gordon}\ and\ \citenamefont
  {Vogelsang}(1993)}]{Gordon:1993qc}%
  \BibitemOpen
  \bibfield  {author} {\bibinfo {author} {\bibfnamefont {L.~E.}\ \bibnamefont
  {Gordon}}\ and\ \bibinfo {author} {\bibfnamefont {W.}~\bibnamefont
  {Vogelsang}},\ }\bibfield  {title} {\enquote {\bibinfo {title} {{Polarized
  and unpolarized prompt photon production beyond the leading order}},}\ }\href
  {\doibase 10.1103/PhysRevD.48.3136} {\bibfield  {journal} {\bibinfo
  {journal} {Phys. Rev. D}\ }\textbf {\bibinfo {volume} {48}},\ \bibinfo
  {pages} {3136} (\bibinfo {year} {1993})}\BibitemShut {NoStop}%
\bibitem [{\citenamefont {Adare}\ \emph
  {et~al.}(2008{\natexlab{b}})\citenamefont {Adare} \emph
  {et~al.}}]{Adare:2008cg}%
  \BibitemOpen
  \bibfield  {author} {\bibinfo {author} {\bibfnamefont {A.}~\bibnamefont
  {Adare}} \emph {et~al.} (\bibinfo {collaboration} {PHENIX Collaboration}),\
  }\bibfield  {title} {\enquote {\bibinfo {title} {{Quantitative Constraints on
  the Opacity of Hot Partonic Matter from Semi-Inclusive Single High Transverse
  Momentum Pion Suppression in Au+Au collisions at $\sqrt{s_{NN}}=200$ GeV}},}\
  }\href {\doibase 10.1103/PhysRevC.77.064907} {\bibfield  {journal} {\bibinfo
  {journal} {Phys. Rev. C}\ }\textbf {\bibinfo {volume} {77}},\ \bibinfo
  {pages} {064907} (\bibinfo {year} {2008}{\natexlab{b}})}\BibitemShut
  {NoStop}%
\bibitem [{\citenamefont {Adare}\ \emph {et~al.}()\citenamefont {Adare} \emph
  {et~al.}}]{PPG212}%
  \BibitemOpen
  \bibfield  {author} {\bibinfo {author} {\bibfnamefont {A.}~\bibnamefont
  {Adare}} \emph {et~al.} (\bibinfo {collaboration} {PHENIX Collaboration}),\
  }\href@noop {} {\enquote {\bibinfo {title} {{Beam energy and centrality
  dependence of direct-photon emission from ultra-relativistic heavy ion
  collisions}},}\ }\bibinfo {note} {{arXiv:1805.04084}}\BibitemShut {NoStop}%
\end{thebibliography}

%
 
\end{document}